\documentclass[prc,aps,twocolumn,floats,floatfix,showpacs,nofootinbib,superscriptaddress]{revtex4-2}

\usepackage{amsmath}
\usepackage{amssymb}
\usepackage{amsfonts}
\usepackage{graphicx}
\usepackage{color}  
\usepackage[english]{babel}
\usepackage{orcidlink}
\usepackage[normalem]{ulem}
\usepackage{soul}

\allowdisplaybreaks

\renewcommand{\vec}[1]{{\mathbf{#1}}}
\newcommand{\nuc}[2]{$^{#1}${#2}}

\newcommand{\vnabla}{\boldsymbol{\mathbf\nabla}}

\newcommand{\br}{\boldsymbol{\mathbf r}}

\newcommand{\nn}{\nonumber}

\begin{document}

\title{Impact of choices for center-of-mass correction energy\\ on the surface energy of Skyrme energy density functionals} 

\author{Philippe~Da~Costa\,\orcidlink{0009-0002-9776-616X}}
\affiliation{Universit\'e Claude Bernard Lyon 1, CNRS/IN2P3, IP2I, UMR 5822,  
rue E.~Fermi, F-69622~Villeurbanne~Cedex, France} 
\author{Karim~Bennaceur\,\orcidlink{0000-0002-6722-491X}}
\affiliation{Universit\'e Claude Bernard Lyon 1, CNRS/IN2P3, IP2I, UMR 5822,  
rue E.~Fermi, F-69622~Villeurbanne~Cedex, France}
\author{Jacques~Meyer\,\orcidlink{0000-0002-7789-4706}}
\affiliation{Universit\'e Claude Bernard Lyon 1, CNRS/IN2P3, IP2I, UMR 5822,  
rue E.~Fermi, F-69622~Villeurbanne~Cedex, France}
\author{Wouter~Ryssens\,\orcidlink{0000-0002-4775-4403}}
\affiliation{Institut d’Astronomie et d’Astrophysique, Universit{\'e} Libre de Bruxelles,
Campus de la Plaine CP 226, 1050 Brussels, Belgium}
\author{Michael~Bender\,\orcidlink{0000-0001-8707-3410}}
\affiliation{Universit\'e Claude Bernard Lyon 1, CNRS/IN2P3, IP2I, UMR 5822,  
rue E.~Fermi, F-69622~Villeurbanne~Cedex, France}

\date{21 March 2023}
%
%

\begin{abstract}

\begin{description}

\item[Background]
In the framework of nuclear energy density functional (EDF) methods, many
nuclear phenomena can be related to the deformation of intrinsic states. Their
accurate modeling relies on the correct description of the change of nuclear
binding energy with deformation. The two most important contributions to the
deformation energy have their origin in shell effects and the surface energy
coefficient of nuclear matter.

\item[Purpose]
It has been pointed out before that the choices made for the 
center-of-mass (c.m.) correction energy and the effective mass
during the parameter adjustment influence the deformation properties 
of nuclear EDFs. We study the impact of these two properties
by means of a set of purpose-built 
parametrizations of the standard Skyrme EDF at next-to-leading (NLO)
order in gradients.

\item[Methods]
In a first step, we build nine series of parametrizations with a
systematically varied surface-energy coefficient $a_{\text{surf}}$ for 
three frequently-used options for the c.m.\ correction (none, one-body term only,
full one-body and two-body contributions) combined with three values
for the isoscalar effective mass $m^*_0/m$  (0.7, 0.8, 0.85) and 
analyse how well each of these parametrizations can be adjusted to 
the properties of spherical nuclei and infinite nuclear matter. 
In a second step, we performed additional fits without the constraint on surface energy, 
adding one ``best-fit'' parametrization to each of the nine series. 
We then benchmark these parametrizations to the deformation properties of heavy nuclei by means of
three-dimensional Hartree-Fock-Bogoliubov calculations that
allow for non-axial and/or non-reflection symmetric configurations.

\item[Results]
We perform a detailed correlation analysis between surface and volume
properties of nuclear matter using the nine series of parametrizations.
The best fits out of each series are then benchmarked on the fission 
barriers of \nuc{240}{Pu} and \nuc{180}{Hg}, as well as on 
the properties of deformed states at normal and superdeformation for 
actinides and nuclei in the neutron-deficient Hg region.

\item[Conclusions]
The main conclusions are as follows:
(i) Each combination of choices for c.m.\ correction and $m^*_0/m$ 
leads to a significantly different optimal value of $a_{\text{surf}}$, reason
being that the effective interaction has to absorb the contribution
of the c.m.\ correction to the total binding energy.
(ii) Many properties of symmetric and asymmetric infinite nuclear 
matter of Skyrme NLO EDFs are strongly correlated to the value 
of $a_{\text{surf}}$.
(iii) Omitting the c.m.\ correction results in values of $a_{\text{surf}}$ that are systematically too small.  
On the other hand, including the one-body term but neglecting the computationally 
expensive two-body term means $a_{\text{surf}}$ will be too large. Both choices result in unrealistic predictions
for fission barriers and superdeformed states of heavy nuclei. Only by incorporating the complete c.m.\ correction
does one obtain quite realistic surface properties from an adjustment protocol that only constrains properties
of infinite nuclear matter and spherical nuclei.
(iv) Lowering $a_{\text{surf}}$ increases the susceptibility of 
finite nuclei to take an exotic shape.

\end{description}

\end{abstract}

\maketitle

%
%
\section{Introduction}
\label{sec:level1}

The self-consistent mean-field approach and its extensions, such as the Random 
Phase Approximation (RPA) and the Generator Coordinate Method (GCM), allow for the 
systematic study of properties and phenomena for all systems
throughout the chart of nuclei~\cite{Bender03r,Schunck19b}.
Using a universal energy density functional (EDF) 
to model the effective in-medium
nucleon-nucleon interaction, these techniques give access to numerous observables 
concerning ground and excited states of nuclei, such as binding energies, 
deformations, isomeric states, rotational bands,
as well as the large-amplitude collective motion of nuclear systems.
Furthermore, symmetry-broken mean-field configurations allow for a natural interpretation of experimental data in terms of the shape of the nucleus in its intrinsic frame.

With the arrival of a wealth of new data on many different aspects of the 
fission process~\cite{Andreyev18r,Schmidt18r} and major advances in its microscopic 
modeling~\cite{Schunck16r,Schunck22r}, there is a renewed interest in 
constructing parametrizations of the nuclear EDFs that are predictive for
physics at large deformation~\cite{Bender20r}.

Indeed, not all parametrizations of the nuclear EDF, most of which are 
mainly adjusted to properties of nuclear matter and finite spherical nuclei, 
describe well the available information on nuclear states at large 
deformation~\cite{Nikolov11a} or fission barriers~\cite{Jodon16a}. There are in 
fact just very few parametrizations of the nuclear EDF that are widely-used for
nuclear fission studies, among which the Skyrme parametrization SkM*~\cite{Bartel82a} 
and the Gogny interaction D1S~\cite{Berger91a} are arguably the most prominent. Both 
were in fact constructed by the readjustment of an earlier parametrization 
that was unable to reproduce even the gross trends of fission barriers.

It is well established that a correct description of shape isomeric states and
fission barriers of heavy nuclei is strongly correlated with the value of the surface 
energy coefficient $a_\text{surf}$~\cite{Bartel82a,Kortelainen12,Jodon16a,Ryssens19a}
and, to a lesser degree, also with the surface symmetry energy coefficient 
$a_\text{ssym}$~\cite{Nikolov11a} of semi-infinite nuclear matter.
There is, however, not a one-to-one correspondence as the actual minima and maxima 
of the deformation energy landscape of finite nuclei are generated by shell effects.
Still, the values of $a_\text{surf}$ and $a_\text{ssym}$ can be indirectly used to 
inform parameter fits about deformation energies~\cite{Bartel82a,Berger91a,Jodon16a}.

The binding energy of finite nuclei is, of course, 
also strongly correlated with $a_\mathrm{surf}$. In a liquid-drop picture of a nucleus with $A$ nucleons,
the surface energy is the only contribution to the binding energy that scales as $A^{2/3}$.
Nuclear masses therefore strongly constrain $a_\mathrm{surf}$ 
even when considering spherical nuclei only.

It has also been pointed out that the choices made to approximately correct for spurious 
center-of-mass (c.m.) motion during the parameter adjustment have an impact on 
the resulting surface properties~\cite{(Ben00f)}. The motivation for such 
correction is that the localized $A$-body states used in the mean-field
modeling of static finite nuclei are not eigenstates of the many-body momentum operator
$\hat{\vec{P}} = \sum_{i} \hat{\mathbf{p}}_i$ with eigenvalue zero, but rather are
superpositions of eigenstates of $\hat{\vec{P}}$ that only yield an average value 
of $\langle \hat{\vec{P}} \rangle = 0$. The resulting spurious excitation
energy can be approximately eliminated by subtracting the average value of 
the kinetic energy of the nucleus in its c.m.\ frame, which is the expectation 
value of the operator~\cite{Stevenson37a,Gartenhaus57a,CHABANAT1998231,(Ben00f)}\footnote{This
expression assumes that one is in the c.m.\ frame of the nuclear system, i.e.\ that
$\langle \hat{\vec{P}} \rangle = 0$. If this is not the case, then the c.m.\ correction
energy is proportional to the dispersion of $\hat{\vec{P}}$, i.e.\
$\frac{1}{2Am} \big( \langle \hat{\vec{P}}^2 \rangle - \langle \hat{\vec{P}} \rangle^2 \big)$, 
instead.}
\begin{equation}
\label{eq:ecm}
\frac{1}{2Am} \hat{\vec{P}}^2
= \sum_{i} \frac{\hat{\mathbf{p}}_i^2}{2Am}
+\sum_{i<j}\frac{\hat{\mathbf{p}}_i \cdot \hat{\mathbf{p}}_j}{Am} \, ,
\end{equation}
where the sums run over occupied single-particle states.
The first term on the 
r.h.s. is a one-body operator that yields  $1/A$ times the free kinetic energy. 
The second term, however, is a two-body operator that leads to a non-local 
contribution to the total energy.

The numerical implementation of the two-body term is comparatively cumbersome, 
and, at least in the context of 
the otherwise local Skyrme EDF, its calculation is quite costly in terms of 
CPU time. For this reason, the two-body term has been omitted for the vast majority of 
parametrizations of Skyrme-type EDFs adjusted so far, including well-known examples such as
SkM*~\cite{Bartel82a},  SLy5s1~\cite{Jodon16a}, SLy4 and SLy5~\cite{CHABANAT1998231}.

Some parametrizations that aim at describing nuclear fusion or fission dynamics were 
adjusted without any c.m.\ correction at all, examples being SLy4d~\cite{Kim97} and 
UNEDF2~\cite{Kortelainen14}. The motivation for the latter practice is that, because of 
the $1/A$ factor, the c.m.\ correction cannot be consistently defined for processes 
where two nuclei fuse or one nucleus splits apart without introducing 
further corrections for other types of spurious motion~\cite{GOEKE1983504,(Ska06)}.

Among the Skyrme parametrizations that were adjusted including the full c.m.\
correction~(\ref{eq:ecm}) are the SkIx and SV-x parametrizations of Refs.~\cite{Rei95,(Klu09)},
SLy6 and SLy7 from Ref.~\cite{CHABANAT1998231}, and those of the large-scale mass fits 
from Refs.~\cite{(Sam02),(Sam04),(Gor07),Scamps21a,Ryssens22a,Ryssens23iib}. The full c.m.\ 
correction is also considered for most of the parametrizations of the Gogny force 
such as D1S~\cite{Berger91a}. 

Disregarding for the moment parametrizations that were explicitly adjusted to
nuclear properties at large deformation, and parametrizations that were 
adjusted with a specific emphasis on other observables than nuclear ground-state 
data, there is a correlation between the scheme for c.m.\ correction and the performance
for fission barrier heights. This observation becomes particularly obvious for parameter sets
constructed within the same protocol, but with different choices for the c.m.\
correction~\cite{(Ben00f)}. Parametrizations that are adjusted with the full c.m.\ correction
give systematically smaller fission barriers than parametrizations that 
keep only the one-body part, but are otherwise adjusted within the same 
fit protocol. This finding is not related to the deformation
dependence of the c.m.\ correction itself, which in general is quite small~\cite{(Ben00f)}. 
Instead, the interaction part of the EDF has to absorb the absent contributions
from Eq.~(\ref{eq:ecm}) to the total binding energy.
There are indications that considering or not 
the c.m.\ correction as such might cause a similar problem: as pointed out in 
Ref.~\cite{Jodon16a}, the SLy4d parametrization~\cite{Kim97} that was adjusted 
with the same protocol as SLy4 and SLy6 but without any c.m.\ correction at all,
gives significantly smaller fission barriers than SLy6.

Among these three families of Skyrme parametrizations, those adjusted with the 
full c.m.\ correction perform systematically better for fission barriers. This 
does not, however, mean that \textit{only} these perform well. Indeed, the long-standing
reference parametrization for fission studies, SkM$^*$~\cite{Bartel82a}, belongs to
the family of parametrizations that only consider the one-body c.m.\ correction.
Similarly, the UNEDF1~\cite{Kortelainen12} and UNEDF2~\cite{Kortelainen14} parametrizations that have been
used in recent fission studies were adjusted without any c.m.\ correction. What these exceptions have
in common is that in one way or the other they were explicitly adjusted to some
characteristics of fission barriers: SkM$^*$ via readjusting some parameters of 
the earlier SkM parametrization~\cite{Krivine80} such that $a_{\text{surf}}$ reproduces a
semi-classical estimate for the fission barrier of \nuc{240}{Pu}~\cite{Bartel82a}, 
whereas the fit protocol of UNEDF1~\cite{Kortelainen12} and UNEDF2~\cite{Kortelainen14} considers excitation energies of some
fission isomers. The adjustment of the D1S parametrization of the Gogny 
force~\cite{Berger91a}, which employs the full c.m.\ correction, was also informed 
by fission barrier heights. Other examples of such parametrizations are those of 
the SLy5sX series that employ only the one-body contribution to the c.m.\ 
correction and which were constructed with a systematically varying constraint
on $a_{\text{surf}}$ with the aim of finding the one that performs best for
fission barriers~\cite{Jodon16a}. That the deformation energy of parameter 
sets adjusted with the full c.m.\ correction is automatically more realistic 
can serve as the starting point for their fine-tuning to fission barriers.
The very recent mass fits BSkG2~\cite{Ryssens22a,Ryssens23iib} and BSkG3 
\cite{grams23a} use the full c.m.\ correction and
achieve a mean deviation of less than 500~keV on the primary and secondary barriers 
of 45~actinide nuclei, including odd and odd-odd ones, through a slight readjustment 
of corrections for other types of collective motion.

Strutinski's theorem~\cite{ragnarsson_nilsson_1995} relates 
deformation energies 
and the actual deformation of energetic minima and barriers to the evolution
of the bunching of single-particle levels around the Fermi energy
with deformation. Deformation properties are therefore also correlated 
to the effective mass~\cite{PhysRevC.76.034317} as the level 
density of single-particle scales with the latter~\cite{MAHAUX19851}.
As a consequence, it has been observed that the effective mass can 
have a visible influence on the excitation energies of superdeformed 
states and fission barrier heights \cite{(Klu09)}.

Starting from these observations, the goals of the present article are 
\begin{enumerate}
\item
to further clarify the correlation between the surface energy coefficient during 
a parameter adjustment and the multiple choices made for the c.m.\ correction 
in the literature;

\item
to further analyze the role of the isoscalar effective mass for fission barriers
and its correlation with the surface energy coefficient.

\end{enumerate}
To this aim, we constructed  new series of parametrizations that are adjusted 
with each of the three different treatments of the c.m.\ correction terms mentioned 
above, and this for three different values of the isoscalar effective mass:
$m_0^*/m = 0.70$, $0.80$ and $0.85$.

This article is organized as follows: Section~\ref{sect:EDF} defines the form
of the Skyrme EDF that will be used for our study, while Sec.~\ref{sect:adjustment}
details the fit protocol used to adjust nine series of new parametrizations 
customized for our study that differ in the scheme for c.m.\ correction and isoscalar
effective mass. Section~\ref{sec:matter} discusses correlations between  
properties of infinite and semi-infinite matter found for these new fits and proposes a set of ``best fits'' for each choice of c.m.\ correction and isoscalar
effective mass that are then used in Sec.~\ref{sect:deformation} for the study 
of representative fission barriers as well as properties of normal-deformed and superdeformed states of heavy nuclei. Section~\ref{sec:summary} summarizes our 
findings.

%
%

\section{The Energy Density Functional}
\label{sect:EDF}

For the purpose of our study of the impact of the scheme for c.m.\ correction and 
the value of the isoscalar effective mass on surface properties of nuclei, we
constructed a set of new parametrizations of the standard Skyrme EDF. As we 
are interested in surface properties, we omit genuine tensor 
forces that directly impact only nuclear shell structure~\cite{Lesinski07,Bender09a} and the response 
to spin- and spin-isospin excitations~\cite{Cao11a}. We also limit ourselves to the 
Skyrme EDF at next-to-leading order (NLO) in gradients~\cite{Raimondi11a,Becker17} and to a form 
where only the coupling constants of the gradientless (leading order) terms 
in the EDF have a (single) density dependence.

The total energy is given by~\cite{Bender03r}
\begin{equation}
\label{eq:Etot}
E_{\mathrm{tot}}
= E_\text{kin} + E_\text{Sky} + E_\text{Cou} + E_{\text{pair}} + E_\text{corr}
  \, ,
\end{equation}
where $E_\text{kin}$ is the kinetic energy, $E_\text{Sky}$ the Skyrme energy that accounts for the binding due to
strong interaction in the particle-hole channel, $E_\text{Cou}$ the Coulomb energy,
$E_{\text{pair}}$ the pairing energy and $E_\text{corr}$ is the sum of all corrections for
quantal zero-point motion.

The kinetic energy is given by~\cite{Bender03r}
\begin{equation}
\label{eq:kin}
E_{\text{kin}} = \frac{\hbar^2}{2m} 
            \, \int \! \mathrm{d}^3r \, \tau_0 (\br) \, ,
\end{equation}
where we use same value $\hbar^2/2m = 20.735530~\text{MeV} \, \text{fm}^2$ for protons and neutrons that is obtained by averaging the values of $\hbar^2/2m_n$ and $\hbar^2/2m_p$ as obtained from the
2020 recommendations for the nucleon masses by the Particle Data 
Group~\cite{PDG2020} and the 2018 CODATA value for $\hbar c$~\cite{CODATA2018}.

The local Skyrme EDF can be decomposed into isoscalar ($t=0$) and isovector ($t=1$) terms
that are either constructed out of time-even (``'')
densities only and terms that contain time-odd (``o'') densities
\begin{align}
E_\text{Sky}
& = \int \! \mathrm{d}^3r \, \sum_{t=0,1} \, 
    \Big[ \mathcal{E}_{t,\text{e}} (\vec{r}) + \mathcal{E}_{t,\text{o}} (\vec{r}) \Big] \, .
\end{align}
We consider here the traditional standard form of the Skyrme EDF for which 
the time-even and time-odd parts take the form~\cite{Bender03r}
\begin{align}
\label{eq:E:sk:e}
\mathcal{E}_{t,\text{e}} (\vec{r})   
& =  C^{\rho \rho}_t  \rho_t^2 (\vec{r})  
   + C^{\rho \rho \rho^\alpha}_t \rho_t^2 (\vec{r})  \, \rho_0^\alpha (\vec{r})  
   \nn  \\   
& + C^{\rho \Delta \rho}_t   \rho_t (\vec{r})\, \Delta \rho_t  (\vec{r}) 
   + C^{\rho \tau}_t            \rho_t (\vec{r}) \, \tau_t  (\vec{r})     
   \nn \\
&  - C^{s T}_t \sum_{\mu, \nu = x}^{z} J_{t,\mu \nu}(\vec{r}) \, J_{t,\mu \nu}(\vec{r})
   \nn  \\   
&   + C^{\rho \nabla J}_t   \, \rho_t (\vec{r}) \vnabla \cdot \vec{J}_t (\vec{r}) \, ,
     \\
\label{eq:E:sk:o}
\mathcal{E}_{t,\text{o}}  (\vec{r})  
& =  C^{ss}_t  \, \vec{s}_t^2  (\vec{r})  
   + C^{ss \rho^\alpha}_t  \, \vec{s}_t^2(\vec{r}) \, \rho_0^\alpha (\vec{r})  
   \nn  \\   
&  + C^{s \Delta s}_t  \vec{s}_t (\vec{r}) \cdot \Delta \vec{s}_t (\vec{r}) 
   - C^{\rho \tau}_t  \vec{j}_t^2 (\vec{r}) 
   \nn  \\   
& + C^{s T}_t  \vec{s}_t (\vec{r}) \cdot \vec{T}_t (\vec{r})
   + C^{\rho \nabla J}_t   \, \vec{s}_t (\vec{r}) \cdot \vnabla \times \vec{j}_t (\vec{r}) \, .
\end{align}
For the definition of the local densities and currents entering the Skyrme EDF see for example Ref.~\cite{Bender03r}.
The coupling constants $C^{\rho\tau}_t$, $C^{sT}_t$, and $C^{\rho \nabla J}_t$ appear in both 
parts of the EDF in order to ensure its Galilean invariance~\cite{Dobaczewski95a}. For the new
parametrizations whose adjustment is described in what follows, the coupling 
constants of the Skyrme EDF are calculated as the strict HF expectation value 
of a central + spin-orbit Skyrme interaction, meaning that the resulting bilinear 
terms in the spin-current tensor density $J_{t,\mu \nu}(\vec{r})$ are kept, as 
is the strict relation $C^{\rho\nabla J}_0 = 3 \,C^{\rho\nabla J}_1$ between the isoscalar 
and isovector spin-orbit coupling constants. In addition, when doing so, the 
coupling constants of all time-odd terms are linearly dependent on the coupling
constants of the time-even terms. Although these relations are necessary to respect 
the Pauli principle (at least for the non-density-dependent
terms), they are not always imposed. Instead, for many parametrizations of Skyrme's EDF, some of the coupling constants are either set to zero or treated as independent ones.
For some of the
existing parametrizations of the Skyrme EDF, some of the time-odd terms in 
Eq.~\eqref{eq:E:sk:o} have to be dropped in order to avoid numerical finite-size 
instabilities~\cite{Les06,Hel13,(Pas13a)}. As will be explained in Sec.~\ref{sect:adjustment}, the 
adjustment protocol for the parametrizations constructed for our study ensures that
none of such instabilities appear at densities that are probed in finite nuclei.

The Coulomb energy of a Slater determinant is given by the sum of a 
direct and an exchange term 
$E_\text{Cou} = E_\text{Cou}^{\text{(d)}} + E_\text{Cou}^{\text{(e)}}$ that 
take the form
\begin{align}
\label{eq:E:cou:d}
E_\text{Cou}^{\text{(d)}}
& = \frac{e^2}{2}
   \iint \, \mathrm{d}^3r_1 \, \mathrm d^3r_2 \,
\frac{\rho_{\text{ch}}(\vec{r}_1) \, \rho_{\text{ch}}(\vec{r}_2)}
     {| \vec{r}_1-\vec{r}_2 |} \, ,
      \\
\label{eq:E:cou:e}
E_\text{Cou}^{\text{(e)}}
& = -\frac{e^2}{4}
        \iint \Big[
         \rho_{\text{ch}}(\vec{r}_1,\vec{r}_2) \, \rho_{\text{ch}}(\vec{r}_2,\vec{r}_1) \nonumber\\
& \quad
 +\vec{s}_{\text{ch}}(\vec{r}_1,\vec{r}_2) \cdot \vec{s}_{\text{ch}}(\vec{r}_2,\vec{r}_1) \Big]
\frac{\mathrm d^3r_1\,\mathrm d^3r_2}{|\vec{r}_1-\vec{r}_2|} \, , 
\end{align}
with $e^2 = 1.439964~\text{MeV}~\text{fm}$ being the square of the 
unit charge~\cite{CODATA2018} and $\rho_\mathrm{ch}$ and $\vec{s}_\mathrm{ch}$
representing, respectively, the scalar and vector charge densities.
As often done for the calculation of the Coulomb energy and fields, we neglect the intrinsic charge 
distribution of nucleons and use point-proton densities instead.

While the direct term only depends on local one-body densities, the exchange
term depends on the full one-body non-local densities $\rho_{p}(\vec{r}_1,\vec{r}_2)$
and $\vec{s}_{p}(\vec{r}_1,\vec{r}_2)$. As we consider only properties of 
doubly-magic nuclei during the parameter adjustment that can be calculated with 
a spherical code in which this term can be treated at acceptable numerical 
cost, the Coulomb exchange energy and its contribution to the mean fields 
are calculated exactly for this task.
When calculating properties of deformed nuclei and fission barriers in 
a Cartesian 3d code, however, the exact numerical treatment of 
$E_{\text{Cou}}^{\text{(e)}}$ becomes unacceptably costly and the 
numerically much more efficient 
Slater approximation that yields a local energy density, 
\begin{equation}
\label{eq:E:cou:e:S}
E_\text{Cou}^{\text{(e,S)}}
= -\frac{3e^2}{4} \left(\frac{3}{\pi}\right)^{1/3}
   \int \! \mathrm d^3r \, \big[ \rho_{\text{ch}}(\vec{r}) \big]^{4/3} \, , 
\end{equation}
is used instead. As analyzed in Refs.~\cite{Skalski01a,Anguiano91a,LeBloas11a}, 
using the Slater approximation introduces only a small error of the order
of 3~\% on the Coulomb exchange energy that only mildly depends on deformation.

For the doubly-magic nuclei entering the parameter adjustment, the HFB
treatment of pairing correlations breaks down such that these calculations 
are performed at the HF level. When calculating deformed open-shell nuclei 
and fission barriers, however, pairing correlations have to be considered. 
The scheme employed for this task will be described in 
Sec.~\ref{sec:mocca_pairing}.

In the present work, $E_\text{corr}$ is limited to the approximate correction for 
the c.m.\ motion and is given by the expectation value of the operator defined in
Eq.~\eqref{eq:ecm}
\begin{equation}
\label{eq:corr}
E_\text{corr}=-E_\text{c.m.}=-E_\text{c.m.}^{(1)}-E_\text{c.m.}^{(2)}
= - \frac{\langle \hat{\vec{P}}^2\rangle}{2m A}\,.
\end{equation}
The c.m.\ correction can be written as the sum of a one-body ($E_\text{c.m.}^{(1)}$)
and a two-body ($E_\text{c.m.}^{(2)}$) contribution, see Eq.~\eqref{eq:ecm}. The 
former is simply proportional to the free kinetic energy
\begin{equation}
\label{eq:cm1}
E_\text{c.m.}^{(1)}
=\frac{E_{\text{kin}}}{A}
=\frac{\hbar^2}{2mA} \int \! \mathrm d^3r \, \tau_0 (\vec{r}) \, ,
\end{equation}
whereas the two-body contribution has to be expressed either through gradients 
acting on the product of non-local densities or as a weighted sum over products of 
off-diagonal matrix elements of the momentum operator, see Ref.~\cite{(Ben00f)}
for the detailed expression. While $E_\text{c.m.}^{(1)}$ is trivial to calculate 
numerically at essentially no cost through Eq.~\eqref{eq:cm1}, the numerical calculation of 
$E_\text{c.m.}^{(2)}$ and the corresponding contribution to the single-particle
Hamiltonian are much more expensive. When working with otherwise local EDFs, 
the two-body c.m.\ correction becomes in fact the single most costly contribution to the energy, 
in particular when self-consistently including its contribution to the single-particle Hamiltonian.

This difference in computational cost, together with the effort necessary 
to implement the comparatively complicated expressions for its contribution 
to the total energy and the single-particle Hamiltonian, are the main 
motivation why $E_\text{c.m.}^{(2)}$ has been omitted for the vast majority 
of parametrizations of Skyrme's EDF, a practice that 
started long ago~\cite{Flugge35a}.

Skyrme's EDF is not the only flavor in use. Both Fayans' EDF~\cite{Fayans98,FAYANS200049}, and the
SeaLL1 EDF of Ref.~\cite{Bulgac18a}  are used and adjusted without any c.m.\ correction at all. 
The Barcelona-Catania-Paris-Madrid (BCPM) EDF~\cite{Baldo08,Baldo10}
employs the analytical estimate of Ref.~\cite{Butler84a} for the
full c.m.\ correction. All of these EDFs have in common that they are local. 
For non-local EDFs that consider the exchange terms from a finite-range force,
there is no computational reason to neglect the two-body part of the 
c.m.\ correction anymore. Consequently, beginning with D1S~\cite{Berger91a}, 
all parametrizations 
of Gogny's force have been adjusted with the full c.m.\ correction, although 
the two-body part is not always used in production calculations~\cite{Batail_thesis}.
Likewise, the parametrizations of the finite-range EDF based on the Michigan-3-Yukawa (M3Y)
force by Nakada~\cite{Nakada10a} as well as the recently introduced regularized finite-range
pseudo-potential~\cite{Dobaczewski_2012,Raimondi_2014,Bennaceur_2017,Bennaceur_2020} also
employ the full c.m.\ correction.

For the parametrizations considering the full c.m.\ correction that we adjusted 
for the present study, we chose a compromise between phenomenology and 
computational cost. 
We treated  $E_{\rm c.m.}^{(2)}$ self-consistently during the parameter adjustments, since doing so is not excessively costly in spherical symmetry and is particularly simple in the absence of pairing.
The Cartesian 3d calculations of deformed nuclei and fission barriers that we describe below, only account for  $E_{\rm c.m.}^{(2)}$ perturbatively for reasons of computational cost. This means in practice that we drop the corresponding contribution to the single-particle Hamiltonian and only add $E_{\rm c.m.}^{(2)}$ to the total energy, in
Eq.~\eqref{eq:Etot}, after convergence. 
Calculating at least part, if not all, of the c.m.\ correction perturbatively is in fact the strategy followed
for many of the existing applications that do consider the full c.m.\ correction~\cite{Friedrich86a,Rei95,(Klu09),Goriely03a,Batail_thesis,Ryssens22a,Ryssens23iib,grams23a}.

We note in passing that the center-of-mass correction approximates the 
energy gain from restoration of translational invariance of 
the nuclear state~\cite{Peierls57a,Peierls62a,Marcos84a,Schmid90a,Rodriguez-Guzman04a,Rodriguez-Guzman04b}, but even including it self-consistently
in the variational equations does by no means even approximatively restore
these symmetries in the wave function. Other observables such as the 
density distributions and its moments therefore also have to be explicitly
corrected for spurious c.m.\ motion as well 
\cite{Tassie58a,Dreizler70a,Schmid90b,Schmid91a}.
For further discussion of the c.m.\ correction to the binding energy and  
its treatment we refer to Refs.~\cite{Bethe37a,Stevenson37a,Elliott55a,Gartenhaus57a,Lipkin58a,Vassell64a,Giraud65a,Aviles68a,Scheid68a,Davies71a,Butler84a,Dietrich96a,(Ben00f)} 
and references therein.

%
%
\section{Parameter adjustment}
\label{sect:adjustment}

%
%
\subsection{General idea}

We have adjusted three sets of parametrizations with different
treatment of the c.m.\ correction. For the first set, we have omitted 
the first and second term of~\eqref{eq:ecm}, such that there is no 
c.m.\ correction at all. These will be labeled by 1F2F(X) in what follows. 
For the second set of parametrizations, the correction was limited to
its one-body part only, i.e.\ the term~\eqref{eq:cm1}. 
These will be labeled by 1T2F(X) in what follows.
Finally, the third set of parametrizations was adjusted considering 
both the one-body and two-body terms in the c.m.\ correction. These
will be labeled by 1T2T(X).
For each choice for the c.m.\ correction we constructed a series of
parameter sets with isoscalar effective mass $m_0^*/m=0.70$, 0.80 and 0.85,
which will be indicated in the parenthesis $(X)$ of the label.
For each of the resulting nine combinations of scheme for c.m.\ correction and 
effective mass, we constructed a series of parametrizations with 
a constraint on the surface energy coefficient $a_{\text{surf}}$ with target 
values varying between 15.5~MeV to 20.0~MeV. For that purpose, $a_{\text{surf}}$ 
is calculated in the computationally-friendly Modified Thomas-Fermi (MTF) 
approximation~\cite{Kri79} that was already used earlier for the same purpose in 
the construction of the SLy5sX parametrizations of Ref.~\cite{Jodon16a}.

%
%

\subsection{Penalty function}
\label{Sect:penalty}

To adjust the coupling constants of the Skyrme EDF of Eqs.~\eqref{eq:E:sk:e} 
and~\eqref{eq:E:sk:o}, we have minimized a penalty function that considers data 
on doubly-magic nuclei and phenomenological properties of infinite nuclear matter (INM).

  This adjustment is achieved by minimizing a penalty function, hereafter
  denoted $\chi^2$, that is a sum of squares of differences between calculated
  quantities ${\cal O}_i$ and their target values ${\cal O}_i^{(0)}$. These are 
  weighted by the inverse of the square of parameters $\Delta{\cal O}_i$ that can
  be regarded as tolerances for the desired final deviation
  \begin{equation}
  \label{eq:penalty}
    \chi^2 
    = \sum_i \left(\frac{{\cal O}_i-{\cal O}_i^{(0)}}{\Delta{\cal O}_i}
             \right)^2 \, .
  \end{equation}
  Reaching the minimum of the objective function  does not guarantee that all 
  quantities ${\cal O}_i$ fall inside of the interval $\big[{\cal O}_i^{(0)}
    -\Delta{\cal O}_i,{\cal O}_i^{(0)}+\Delta{\cal O}_i\big]$.

  Two quantities characterizing infinite nuclear matter are not constrained
  through the minimization of the penalty function~\eqref{eq:penalty}, but 
  are enforced to take a definite value. These quantities are the saturation density, 
  which is fixed to $\rho_{\text{sat}} = 0.16~\text{fm}^{-3}$, and the isoscalar 
  effective mass $m^*_0/m$ that is set to the required value for each series
  of fits. Since the objective function involves several properties of 
  nuclear matter at saturation, fixing $\rho_{\text{sat}}$ in this way greatly stabilizes 
  the parameter adjustment.

Adapting the protocol used for adjusting parametrizations of 
Refs.~\cite{Jodon16a}, the set of constraints considered here is the following:
\begin{itemize}
\item 
Total energies of seven doubly-magic nuclei from AME20~\cite{Wang_2021},  listed in Table~\ref{tab:nuc}.

\item 
The difference in binding energy $\Delta E$ between $^{56}$Ni and $^{40}$Ca
depends strongly on the distance between the neutron $1f_{7/2}$ and
$1d_{3/2}$ orbitals and therefore on the strength of the spin-orbit term.
We have put a constraint on $\Delta E$ with a target value of $141.920$~MeV
and a tolerance of 1~MeV to constrain this term.

\item Properties of symmetric infinite nuclear matter in the vicinity of the 
saturation point: energy per nucleon $\varepsilon_{\text{sat}}$, symmetry energy coefficient
$J$ and its slope $L$ with target values and tolerances given on Table~\ref{tab:inm}.
The choice of target values for $J$ and $L$ is motivated by microscopic calculations
in infinite nuclear matter~\cite{Lattimer23a}.

\item Energy per nucleon in infinite neutron matter. We used values
calculated for the potentials UV14 plus UVII (see Table~III in~\cite{(Wir88)})
at densities up to 0.45~fm$^{-3}$ with a tolerance of 25~\%.

\item Energy per nucleon in polarized infinite nuclear matter and neutron
matter. Adjustment of parameters sometimes leads to the appearance of a bound
state in symmetric polarized matter or to the collapse of polarized neutron
matter at high density. To avoid this type of results, we used the constraints
of $E/A = 12.52~$MeV at density 0.1~fm$^{-3}$ in polarized nuclear matter
and $E/A = 40.10~$MeV in polarized neutron matter at the same density
(taken from Ref.~\cite{Bordbar08a}) both with a large tolerance of 25~\%.

\item 
To avoid the appearance of finite-size instabilities~\cite{(Pas13a)}
we used the linear response method~\cite{Hel13} to enforce that the lowest
poles of the response function remain above $\rho_\mathrm{min} = 1.2 \times \rho_{{\text{sat}}}
\simeq 0.192~\text{fm}^{-3}$ in symmetric nuclear matter for all spin and
isospin channels (except for the case of the spinodal instability at low
density in the $(S,T)=(0,0)$ channel) and above half of this density in
pure neutron matter, see Ref.~\cite{(Pas13a)}.
An instability is characterized by a divergence of the
  response function, or a zero of its inverse, for given values of
  the density $\rho_0$ and the transferred momentum $q$ between particles and
  holes. To push any instability above $\rho_\mathrm{min}$, we
  calculate the sum of the modulus of the inverse of the response function
  at $\rho_0=\rho_\mathrm{min}$ for equally spaced values of $q$ from
  0 to 9~fm$^{-1}$ with $\delta q=0.01~\mathrm{fm}^{-1}$
  and require the result, in all $(S,T)$ channels to
  be greater than 0 using an asymmetrical constraint as described by
  equation~(58) in~\cite{Bennaceur_2017}.
These strong constraints allow us to avoid the appearance
of finite-size instabilities for all parametrizations constructed, as we checked explicitly afterwards.

\item 
The surface energy coefficient calculated in semi-infinite nuclear matter using the MTF
approximation was constrained to a series of values in steps of 0.2~MeV with a tolerance of 0.01~MeV.
Nine series of interactions have been thus constructed labeled with their \{center of mass, isoscalar effective mass, MTF surface energy coefficient\} options.
Beyond these ones, nine other interactions (only labeled with their \{center of mass, isoscalar effective mass\} options) have been built omitting the constraint of the surface coefficient
in order to have in each case the exact minimum of the penalty function.

\end{itemize}

The power $\alpha$ of 
the density dependence in Eq.~\eqref{eq:E:sk:e} is not considered as a free parameter
but set to $\alpha = 1/6$. It is well known that with standard NLO Skyrme functionals
and for given values of saturation density and energy per nucleon in symmetric infinite
nuclear matter, the isoscalar effective mass $m^*_0/m$ and the compression modulus $K_\infty$
are not independent quantities~\cite{CHABANAT1997710}. The choice to set $\alpha$ to $1/6$ allows to vary the
isoscalar effective mass while keeping $K_\infty$ in a acceptable interval, see Sec.~\ref{sec:bulk}.
With this choice for $\alpha$ and for the chosen fixed values for $\rho_\mathrm{sat}$ and $m_0^*/m$,
the EDF contains in total seven free parameters that have to be adjusted.

\begin{table}[t!]
\caption{\label{tab:nuc}
Binding energies (in MeV) of doubly-magic nuclei used to constrain the
parameters of the functional. Note that the value for $^{78}$Ni
is extrapolated. The last column gives the tolerances (in MeV) used in the fit protocol.
}
\begin{tabular}{rrr}
\hline\noalign{\smallskip}
  Nucleus & \multicolumn{1}{c}{$E_\mathrm{tot}$}  &
               \multicolumn{1}{c}{Tolerance}  \\ 
\noalign{\smallskip}\hline\noalign{\smallskip}
  $^{40}\text{Ca}$   & $-342.034$   &  $\pm 1.0 $   \\
  $^{48}\text{Ca}$   & $-415.983$   &  $\pm 1.0 $   \\
  $^{56}\text{Ni}$   & $-483.954$   &  $\pm 1.0 $   \\
  $^{78}\text{Ni}$ & $-642.522$   &  $\pm 2.0 $   \\
  $^{100}\text{Sn}$  & $-824.995$   &  $\pm 1.0 $   \\
  $^{132}\text{Sn}$  & $-1102.675$  &  $\pm 1.0 $   \\
  $^{208}\text{Pb}$  & $-1635.862$  &  $\pm 1.0 $   \\
\noalign{\smallskip}\hline
\end{tabular}
\end{table}

\begin{table}[t!]
\caption{
Properties of symmetric nuclear matter in the vicinity of saturation used to constrain the EDF parameters.
All quantities are in~MeV.\label{tab:inm}
}
\begin{tabular}{lcc}
\hline\noalign{\smallskip}
  Property & Target value & Tolerance  \\ 
\noalign{\smallskip}\hline\noalign{\smallskip}
  $\varepsilon_{\text{sat}}$          & $-16.0$        & $\pm 0.1$ \\
  $J  $          &  32            & $\pm 1$     \\
  $L  $          &  50            & $\pm 5$     \\
\noalign{\smallskip}\hline
\end{tabular}
\end{table}

All nuclei considered for the fit of parameters are doubly-magic spherical 
nuclei. It is assumed that pairing correlations do not contribute to those, 
such that calculations are done at the HF approximation. Their numerical calculations 
was performed on a radial mesh in coordinate space with 80~points with 
a constant spacing of $0.25$~fm using the code FINRES$_4$~\cite{FINRES}.

%
%
\section{Correlations between nuclear matter properties}
\label{sec:matter}

%
%
\subsection{From EDF parametrization to Liquid-drop model}
\label{sec:LDM}

As has been pointed out before~\cite{(Ben00f),Nikolov11a}, the physical origin 
of the correlation between choices made for the center-of-mass correction 
and nuclear surface properties on the one hand, and of the correlations 
between surface properties and the bulk properties of nuclear matter on 
the other hand, can be understood when looking at binding energies obtained 
from a liquid-drop model (LDM) whose parameters are set to the values 
predicted by the paramerizations of EDF models.

To this aim, we employ the following form for the LDM energy of 
a nucleus with $N$ neutrons and $Z$ protons 
\begin{align}
\label{eq:mac}
E_{\text{LDM}} (N,Z)
& = ( a_{\text{vol}}  + a_{\text{sym}} \, I^2 ) \, A
      \nonumber \\
&   \quad
      + ( a_{\text{surf}} + a_{\text{ssym}} \, I^2 ) \, A^{2/3}
      \nonumber \\
&    \quad 
      + \frac{3 \, e^2}{5 \, r_0} \, \frac{Z^2}{A^{1/3}}
      - \frac{3 \, e^2}{4 \, r_0} \left( \frac{3}{2\pi}\right)^{2/3} \,
        \frac{Z^{4/3}}{A^{1/3}} \, ,
      \nonumber \\
&   
\end{align}
where \mbox{$A = N+Z$} is the mass number and \mbox{$I=\frac{N-Z}{N+Z}$} the
isospin asymmetry. 
The coefficients of the volume ($a_{\text{vol}}$) and volume symmetry 
($a_{\text{sym}}$) energy can be related to properties of infinite
nuclear matter at the saturation point, whereas the coefficients of the 
surface ($a_{\text{surf}}$) and surface symmetry ($a_{\text{ssym}}$) energy 
are connected to properties of semi-infinite nuclear matter (SINM). The radius constant $r_0$ 
is determined by the nuclear matter saturation
density $\rho_{{\text{sat}}}$ through $r_0^3 = 3/(4 \pi \rho_{\text{sat}})$. 
Constructing an LDM that accurately approximates the binding energies of a self-consistent model
   would require additional terms~\cite{(Rei06)}, but the simplicity of Eq.~\eqref{eq:mac} is sufficient for our study.

%
%
\subsection{Further analysis of the correlations between the values for 
            $a_{\text{surf}}$ obtained through different schemes}
\label{sect:asurf}

We recall that there are several approaches to calculate the surface and surface
symmetry energy coefficients of an EDF that differ in their strategy and 
computational cost. A widely used procedure is to extract $a_{\text{surf}}$
and $a_{\text{ssym}}$ from calculations of the model system of SINM~\cite{Cote78,Jodon16a}
\begin{align}
\label{eq:asurfeff}
a_{\text{surf,eff}}(I)
& = a_{\text{surf}} + a_{\text{ssym}} \, I^2
     \nn \\
& =  \lim_{L \to \infty} \Bigg\{ \bigg[ 4 \pi r_0^2 \int_{-L/2}^{+L/2} \! \mathrm{d}z 
      \, \mathcal{E}(z) \bigg]
      - E_{\text{ref}}(I,L) \Bigg\} 
\end{align}
in a one-dimensional box of length $L$, where $\mathcal{E}(z)$ is the 
energy density of SINM calculated at an asymmetry $I$ and $E_{\text{ref}}(I)$ is a reference volume energy
that depends on nucleon numbers.
 
The value of $a_{\text{surf}}$ can be determined from a single SINM 
calculation of symmetric matter ($I=0$), whereas the extraction of 
$a_{\text{ssym}}$ requires at least two calculations at different
asymmetries $I$.

The SINM calculations can either be performed in some variant of the 
semi-classical Thomas-Fermi approximation or in the quantal Hartree-Fock framework.
While each of these schemes yields a slightly different value for 
$a_{\text{surf}}$, it has been argued in Ref.~\cite{Jodon16a} that they
are basically equivalent for the purpose of constraining effective 
interactions as long as the value that $a_{\text{surf}}$ is constrained 
to is suitably chosen.

To put the earlier analysis of Ref.~\cite{Jodon16a} onto a wider basis of 
parametrizations that systematically cover a wide interval of $a_{\text{surf}}$ 
values, and to confirm that the main conclusions of this study also apply when 
making the choices of 
the fit protocol described above, we compare values for $a_{\text{surf}}$
extracted from calculations of semi-infinite nuclear matter performed within
either the Hartree-Fock (HF), within the semi-classical Extended Thomas-Fermi (ETF) 
approach up to order $\hbar^4$, or within the Modified Thomas-Fermi (MTF) approach. 
For details about these methods we refer to Ref.~\cite{Jodon16a} and references 
therein, and recall only their main characteristics. 
In a quantal HF calculation of SINM, one minimizes the total
energy as calculated from the  self-consistent densities of a 
Slater determinant of single-particle states. In the ETF calculation, the 
kinetic and spin-current densities entering the Skyrme EDF are developed into  
functionals of the local density and its derivatives. The surface energy is 
then minimized with respect to the parameters of a prescribed profile for the 
local densities of protons and neutrons. Finally, the MTF approach is based 
on the observation that a slight modification of the relative weights of the 
semi-classical expansion of the kinetic density in a limited ETF expansion
up to order $\hbar^2$ makes the 
system integrable for standard Skyrme EDFs at NLO~\cite{Kri79,Treiner86a}, 
such that the optimal density profile is obtained without a variational
calculation~\cite{Treiner86a,Jodon16a}. The computational cost is thereby
considerably reduced when going from HF to ETF and then to MTF.

The semi-classical calculations reported here were performed with the same 
tools as those reported in Ref.~\cite{Jodon16a}, whereas the HF calculations were 
made with a newly designed code~\cite{AtreusSINM} that yields results for 
$a_{\text{surf}}$ that are identical to those reported in Ref.~\cite{Jodon16a}
within typically 0.01~MeV.

For the surface symmetry energy coefficient $a_{\text{ssym}}$, the 
discussion will be limited to values extracted from HF calculations of 
SINM. The reason is that there are several different 
choices for the reference energy $E_{\text{ref}} (I,L)$ entering 
Eq.~\eqref{eq:asurfeff} that are frequently used in the literature for 
its extraction and that lead to different values of $a_{\text{ssym}}$ 
when extracted from the same calculations of semi-infinite
matter. As $a_{\text{ssym}}$ has not been constrained during the 
adjustment of the parameter sets discussed here, we will limit 
its analysis to one scheme to calculate $a_{\text{ssym}}$ and to 
one procedure to extract it. For the latter, we choose the 
thermodynamical definition~\cite{Myers85a,Farine86a}, where the
reference energy $E_{\text{ref}} 
= \epsilon_{\text{F},n} \, N_{\text{box}} + \epsilon_{\text{F},p} \, Z_{\text{box}}$
is provided by the Fermi energies of protons and neutrons, respectively,
and the number of protons and neutrons that enter the calculation of the
energy density $\mathcal{E}(z)$ in Eq.~\eqref{eq:asurfeff}.

\begin{figure}[t!]
        \includegraphics{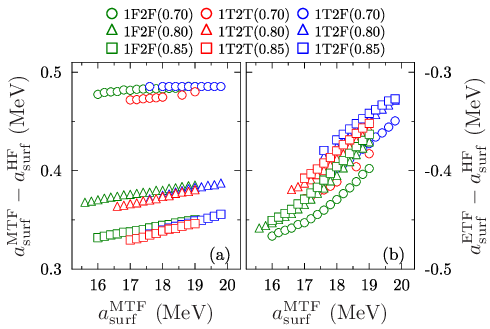}
        \caption{\label{fig:asurf:correlation}
        Differences between the surface energy coefficient as 
        calculated in either the MTF (panel (a)) or ETF (panel (b))
        approach and its HF value for the nine series of parametrizations as indicated as a function of their 
        $a_{\text{surf}}^{\text{MTF}}$.
        Different colors indicate different schemes for c.m.\
        correction, whereas different markers indicate different isoscalar 
        effective mass $m^*_0/m$.
        }
\end{figure}

Figure~\ref{fig:asurf:correlation} shows the differences
$\Delta a_{\text{surf}}^{\text{MTF}} = a_{\text{surf}}^{\text{MTF}} - a_{\text{surf}}^{\text{HF}}$ 
and 
$\Delta a_{\text{surf}}^{\text{ETF}} = a_{\text{surf}}^{\text{ETF}} - a_{\text{surf}}^{\text{HF}}$ 
between the values of the surface energy coefficient extracted from SINM
calculations with either of the semi-classical ETF and MTF approaches and its value obtained 
from a HF calculation for all nine series  of fits by systematically varying
$a_{\text{surf}}^{\text{MTF}}$ and the isoscalar mass $m^*_0/m$.

The difference $\Delta a_{\text{surf}}^{\text{MTF}}$ between MTF and HF results shows a clear 
dependence on $m^*_0/m$, which was already hinted in the results discussed in Ref.~\cite{Jodon16a}: 
the difference decreases with increasing effective mass. This can be explained by the nature of the 
MTF approximation that modifies the dependence of the kinetic density $\tau[\rho(\vec{r})]$
on the local density $\rho(\vec{r})$ in the semi-classical approximation in such a way that the problem becomes integrable~\cite{Kri79,Jodon16a}:
the smaller the 
isoscalar effective mass, the larger becomes the relative contribution from EDF
terms that contain products of $\tau(\vec{r})$ and other densities compared to the 
free kinetic energy that is linear in $\tau(\vec{r})$. It appears that the 
MTF approximation works better for terms of the latter type than those of the former.

For each fixed value of $m^*_0/m$, there also is a very mild dependence of 
$\Delta a_{\text{surf}}^{\text{MTF}}$ on the actual value of $a_{\text{surf}}^{\text{MTF}}$.

For comparison, and to complement the discussion of Ref.~\cite{Jodon16a}, 
Fig.~\ref{fig:asurf:correlation} also shows the difference 
$\Delta a_{\text{surf}}^{\text{ETF}}$ between ETF and HF results.
Although the deviation is not the same for all, its spread is much smaller,
meaning that the ETF approximation works much more consistently for 
parametrizations with different $m^*_0/m$. The difference, however, depends
more strongly on the value of $a_{\text{surf}}^{\text{MTF}}$ that the 
parameter set is constrained to than it is the case for $\Delta 
a_{\text{surf}}^{\text{MTF}}$.
At fixed $a_{\text{surf}}^{\text{MTF}}$ the deviation is systematically slightly 
smaller for parametrizations with larger $m^*_0/m$. There also is a slight 
systematic dependence of $\Delta a_{\text{surf}}^{\text{ETF}}$ on the scheme 
used for the c.m.\ correction, where the deviation is smallest for parameter 
sets adjusted with the full 1T2F scheme and largest for those adjusted 
with the 1F2F scheme.

%
%
\subsection{Penalty function and correlations between nuclear matter properties}
\label{sect:penalty:syst}
%
%
\subsubsection{Surface properties}

\begin{figure}[t!]
        \includegraphics{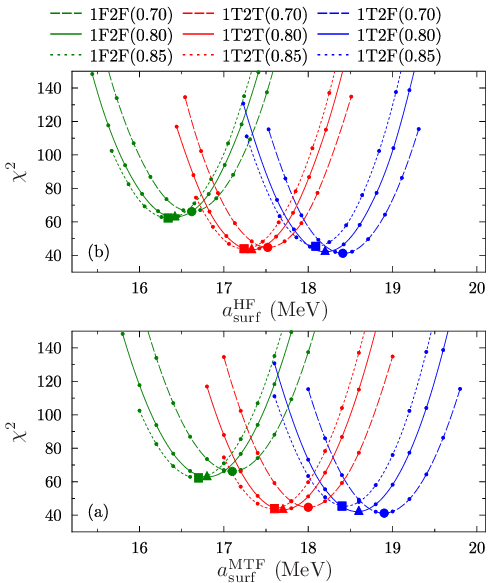}
        \caption{\label{chi2}
        Penalty function $\chi^2$ for all parametrizations out of the nine series 
        of fits as a function of their value of $a_\text{surf}^\text{MTF}$ 
        (panel (a)) and $a_\text{surf}^\text{HF}$ (panel (b)).
        The dots indicate the parametrizations with systematically varied 
        $a_\text{surf}^\text{MTF}$ whose properties are studied in 
        Sec.~\ref{sect:penalty:syst}.
        For each of the nine series, the minimum of $\chi^2$ is indicated by a filled 
        marker. The properties of the corresponding ``best fit'' 
        parametrizations are studied beginning with Sec.~\ref{sect:penalty:best}.
     }
\end{figure}

Figure~\ref{chi2} shows the penalty function for all parametrizations
out of the nine series of fits with systematically varied $a_{\text{surf}}^{\text{MTF}}$ 
and $m^*_0/m$. We focus first on panel (a) that shows the penalty
function as a function of $a_{\text{surf}}^{\text{MTF}}$.

There is a clear correlation between the value of $a_\text{surf}^{\text{MTF}}$ 
at the minimum of the penalty function and the scheme of c.m.\ correction:
for parameter sets using the popular 1T2F scheme, the minimum is can be found near 18.6~MeV,
while for parameter sets using the full 1T2T scheme it is located around 17.7~MeV, and for 
parameter sets without any c.m.\ correction (1F2F)
one finds it at 16.8~MeV. The position of the minimum also depends in a more limited 
way on the effective mass $m^*_0/m$: with increasing effective mass, the minimum shifts
to smaller values of $a_\text{surf}^{\text{MTF}}$.
The different locations of the minima in terms of the surface energy of the parametrizations
are quite meaningful: a difference of 1~MeV in $a_\text{surf}$ 
typically changes the outer fission barrier heights of actinide nuclei
by about 4~MeV, a value that is comparable to the experimentally determined barriers for these nuclei.
We illustrate this in Sec.~\ref{sect:deformation} but it was already pointed out repeatedly in earlier studies~\cite{Dutta80,Brack85,Jodon16a,Ryssens19a}.

Previous studies conducted in Refs.~\cite{Jodon16a,Ryssens19a} have shown that, 
for nuclear EDFs that do not consider other quantal corrections\footnote{In 
the presence of additional strongly deformation dependent quantal corrections
such as a rotational correction~\cite{Gogny_D1,Berger89_D1S,Ryssens23iib}, 
or when considering exact restoration of angular momentum~\cite{Bender04a,Marevic20a}, 
the optimal value of $a_\text{surf}^\text{MTF}$ can be substantially different
as it only represents the deformation dependence of the interaction energy, 
but not the deformation dependence of the quantal corrections.
}
than possibly a c.m.\ correction as done here, the optimal values of 
$a_\text{surf}^\text{MTF}$ for a satisfying description of the 
deformation properties  of heavy nuclei fall into an interval 
between 17.6 and 18.0~MeV. It is striking to see that 
the minimum of the penalty function $\chi^2$ as a function of $a^{\text{MTF}}_\text{surf}$ 
is situated precisely in this interval for the parametrizations of 
the 1T2T type, whereas it is well above for parametrizations of 1T2F type
and well below for parametrizations of 1F2F type. 

The systematic differences between the values of $a_\text{surf}^\text{MTF}$ 
at the minima of the penalty function explain why many of the existing 
parametrizations that use the popular 1T2F recipe systematically fail to 
describe fission barrier heights and grossly overestimate 
them~\cite{CHABANAT1998231,(Ben00f),Jodon16a}, 
unless their surface properties 
are constrained during the parameter adjustment. One representative example is the SLy4d parametrization~\cite{Kim97},
whose surface properties and fission barriers where discussed in Ref.~\cite{Jodon16a}.

It is possible to constrain 1F2F and 1T2F parametrizations to realistic surface
properties during the fit, but this comes at the price of a deteriorated description
of other observables that enter the penalty function. 
For example, the SLy5s1 parametrization
has $a_\text{surf}^\text{MTF} = 18.0~\text{MeV}$, but performs comparatively poorly
for binding energies of nuclei~\cite{Ryssens19a}.
Recalling that SLy5s1 is of 1T2F(0.70) type and has been adjusted 
with a protocol that is almost identical to ours, this finding can be easily 
understood from panel (a) of Fig.~\ref{chi2}.
Bringing $a_\text{surf}^\text{MTF}$ to a realistic value is only achieved 
at the expense of other features.

As illustrated by Fig.~\ref{fig:asurf:correlation}, the offset between the
HF and MTF values for $a_\text{surf}$ slightly depends on the isoscalar 
effective mass of the parametrizations. The change of this offset has as a 
consequence that, for a given choice of scheme for c.m.\ correction, the 
minima of the penalty function $\chi^2$ for different choices of isoscalar 
effective mass $m^*_0/m$ become closer when plotting $\chi^2$ as a function 
of $a_\text{surf}^{\text{HF}}$ instead of
the value of $a_\text{surf}^{\text{MTF}}$ that the parametrizations were 
constrained to, see panel (b) of Fig.~\ref{chi2}.
This observation indicates that the fission barrier
heights of optimal fits that employ the same scheme for c.m.\ correction might 
depend less on the effective mass than is apparent on panel (a) of Fig.~\ref{chi2}, 
at least if one assumes that $a^{\rm HF}_{\rm surf}$ is the value of the surface energy that is the most 
directly correlated to the deformation energies obtained in EDF calculations.

\begin{figure}[t!]
        \includegraphics{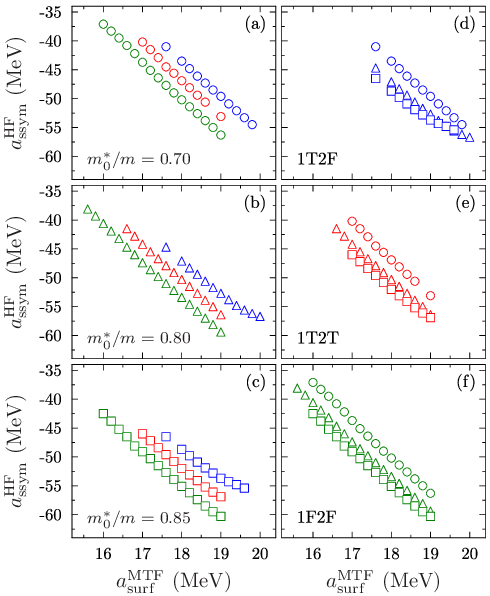}
        \caption{\label{fig:asurfsym}
        Surface symmetry energy coefficient $a_{\text{ssym}}^{\text{HF}}$ calculated 
        in HF approximation for all parametrizations out of the nine series of 
        fits, plotted as a function of their value for $a_{\text{surf}}^{\text{MTF}}$.
        Panels (a), (b), and (c) compare parametrizations with same effective mass but
        different scheme for c.m.\ correction, whereas panels (d), (e), and (f) compare
        parametrizations with same scheme for c.m.\ correction but different effective
        mass $m^*_0/m$.
        }
\end{figure}

For a given nucleus, the effective surface energy coefficient $a_{\text{surf,eff}}(I)$
of Eq.~\eqref{eq:asurfeff} also depends on its asymmetry $I$ through the surface symmetry energy coefficient $a_{\text{ssym}}$.
As a consequence, the correlation between the fission barrier of this nucleus and the value of $a_{\text{surf}}$ constrained
in a parameter fit also depends on the value adopted by $a_{\text{ssym}}$ during the parameter fit. As it turns out, for our
series of parameter fits the values of $a_{\text{ssym}}$ are not identical, but also correlated to the value of $a_{\text{surf}}^{\text{MTF}}$,
the choice of scheme for c.m.\ correction, and the isoscalar effective mass as illustrated on Fig.~\ref{fig:asurfsym}.

Within each series of our fits, the absolute value of $a_{\text{ssym}}$ increases 
with $a_{\text{surf}}$. As $a_{\text{ssym}}$ and $a_{\text{surf}}$ have in general 
opposite sign, this dependence keeps the values of $a_{\text{surf,eff}}(I)$ of 
very asymmetric nuclei closer together when comparing different parametrizations 
out of a given series than their difference in $a_{\text{surf}}$ would suggest.

For a given scheme for c.m.\ correction, the absolute value of $a_{\text{ssym}}$ 
increases with effective mass, typically by about 4~MeV
when going from $m^*_0/m = 0.7$ to $m^*_0/m = 0.8$, and by about another 
2~MeV when going from $m^*_0/m = 0.80$ to $m^*_0/m = 0.85$.
For a given effective mass, the absolute value of $a_{\text{ssym}}$ increases
by about 3~MeV when going from the 1T2F scheme to the 1T2T
scheme, and by about another 3~MeV when going from the 1T2T 
scheme to the 1F2F scheme.

We note in passing that we made an unsuccessful attempt to simultaneously 
constrain $a_{\text{surf}}$ and $a_{\text{ssym}}$ at the MTF level, using an adaptation of the estimate of $a_{\text{ssym}}$ proposed in Ref.~\cite{Krivine83}.
As it turns out, this cannot be meaningfully done for the NLO Skyrme EDF 
that we use here:
setting $a_{\text{ssym}}$ to a value that differs significantly 
from the optimal value for given $a_{\text{surf}}$ indicated by 
Fig.~\ref{fig:asurfsym} pushes the values of some nuclear matter properties far out 
of their accepted range.
Adding $a_{\text{ssym}}$ to a fit protocol that already fixes
$a_{\text{surf}}$ and aims at realistic bulk properties over-constrains
the parameter fit of a Skyrme EDF at NLO. This is not surprising
in view of the limited number of independent coupling constants of
Eq.~\eqref{eq:E:sk:e} that determine nuclear matter properties.
It remains to be shown if more general forms of a Skyrme EDF would allow for 
a fine-tuning of $a_{\text{ssym}}$ without deteriorating the 
bulk properties.

%
%
\subsubsection{Bulk properties}
\label{sec:bulk}

\begin{figure}[t!]
        \includegraphics{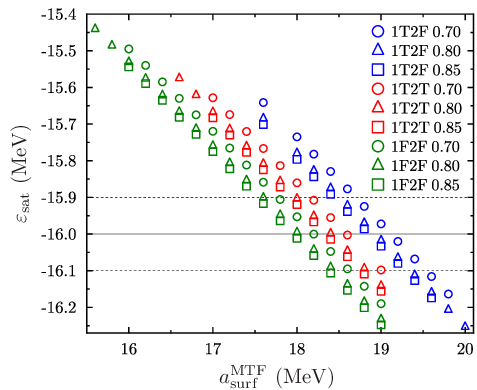}        
        \caption{\label{fig:asurf:ebya}
        Location of the saturation point:
        Energy per particle $\varepsilon_{\text{sat}}$ at the saturation density 
        $\rho_{\text{sat}}$ of the parametrizations as indicated.
        The horizontal lines indicate the target value of $\varepsilon_{\text{sat}}$
        and its tolerance in the penalty function. 
        }
\end{figure}

\begin{figure}[t!]
\includegraphics{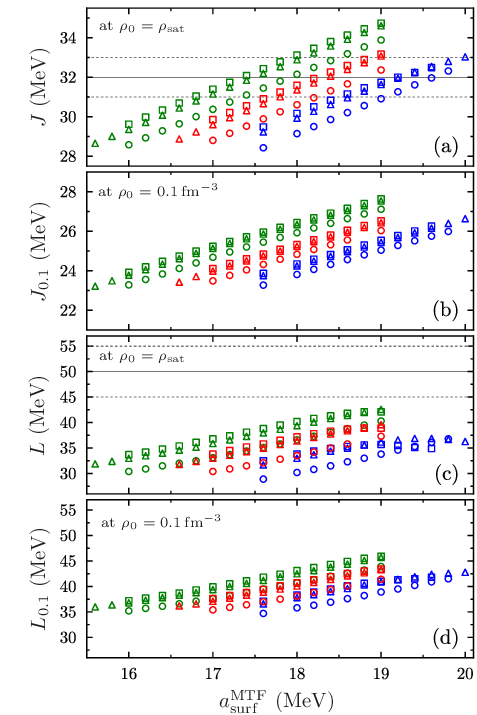}
        \caption{\label{fig:asurf:JL}
        Symmetry energy $J$ and its slope $L$
        at saturation density $\rho_{{\text{sat}}}$ and  
        at $\rho_0 = 0.1 \, \text{fm}^{-3}$ (see text)
        for the parametrizations as indicated.
        The horizontal lines indicate the target values of $J$ and $L$
        in the penalty function, as well as their 
        respective tolerances. Colors and symbols as in 
        Fig.~\ref{fig:asurf:ebya}.}
\end{figure}

As it turns out, for a Skyrme EDF at NLO the infinite nuclear 
matter properties are already strongly correlated to the value of 
$a_{\text{surf}}$ even when $a_{\text{ssym}}$ is left unconstrained. 

Figure~\ref{fig:asurf:ebya} displays the energy per particle $\varepsilon_{\text{sat}}$
at saturation density of homogeneous INM.
This property equals the volume energy coefficient $a_{\text{vol}} = \varepsilon_{\text{sat}}$ 
of the liquid drop model. We constrain it in our parameter fits to
$\varepsilon_{\text{sat}} = (16.0 \pm 0.1)~\text{MeV}$, although this target value is incompatible 
with the most extreme values of $a_{\text{surf}}$ 
covered by our fits. In general, large (positive) values of $a_{\text{surf}}$ correspond
to large negative values of $a_{\text{vol}}$ and vice versa. This correlation
can be understood when considering the role of the surface energy for 
nuclear masses in the liquid-drop model. Since the surface energy 
$a_{\text{surf}} \, A^{2/3}$ reduces nuclear binding, the coefficients of other terms in the 
liquid-drop model have to change in a way that increases their contribution
to the total energy in order to keep binding energies 
of finite nuclei roughly constant. The volume term is apparently one of them. 
Not surprisingly, $\varepsilon_{\text{sat}}$ is close to $16.0~\text{MeV}$ 
for parametrizations with a value of $a_{\text{surf}}^{\text{MTF}}$ 
near the minimum of the penalty function for all nine series of fits.

Because of the different $A$ (and $I$) dependence of their contribution to 
total binding energy, one term can of course not perfectly compensate for 
the change of the other, such that multiple nuclear matter properties change 
when varying $a_{\text{surf}}^{\text{MTF}}$. And indeed, as can be seen from 
Fig.~\ref{fig:asurf:JL}, the volume symmetry energy coefficient 
$a_{\text{sym}}$ of the liquid-drop model, which equals the symmetry 
energy of symmetric matter
\begin{equation}
S(\rho_0) = \frac{1}{2} \, \frac{\partial^2}{\partial I^2}\,\frac{\mathcal{E}}{\rho_0}
\end{equation}
at saturation density, i.e.
\begin{equation}
    a_{\text{sym}} = J = S(\rho_{\text{sat}})\,,
\end{equation}
is also evolving with the constrained value of $a_{\text{surf}}^{\text{MTF}}$
over a wide range between roughly 28.5 and 35~MeV. Within
each series of fits, $J$ almost linearly increases with $a_{\text{surf}}^{\text{MTF}}$. While the slope of this dependence 
is almost the same within all nine sets of fits, there is a large offset
between different series that strongly depends on the choice made for the 
scheme for c.m.\ correction (indicated by different symbols in 
Fig.~\ref{fig:asurf:JL}) and to a much lesser degree also on the value 
for the effective mass $m_0^*/m$  (indicated by different colors). 

In addition, Fig.~\ref{fig:asurf:JL} displays the slope 
of the symmetry energy at the saturation point
\begin{equation}
L = \left.3 \rho_0 \, \frac{\partial S(\rho_0)}{\partial \rho_0}\right|_{\rho_0=\rho_\mathrm{sat}} \, ,
\end{equation}
as well as the values of the symmetry energy and its slope at $\rho = 0.1~\text{fm}^{-3}$
denoted $J_{0.1}$ and $L_{0.1}$. It has 
been pointed out that the slope $L$ is correlated with characteristics
of asymmetric nuclear systems at densities that are very different 
from saturation density. Examples for such systems are finite nuclei with neutron skins,
heavy-ions in collision, or neutron stars~\cite{Steiner05a,Horowitz14a,Baldo16a,RocaMaza18a,Lattimer23a}.

One can observe that the values obtained for $L$ are well outside of the
interval defined by the target value and the tolerance $[45,55]$
(in MeV). This feature reveals that the EDF~\eqref{eq:E:sk:o} does not
have the required flexibility to satisfy all constraints listed in
Sec.~\ref{Sect:penalty} within the tolerance intervals. Even
if the values obtained for $L$ are rather low to correctly reproduce
global properties of neutron stars, we can consider that this will
not impact too much the properties of finite nuclei besides, possibly,
neutron skins, since these values are close to the one obtained with
other successful interaction such as D1S~\cite{Berger89_D1S}.

Besides the correlations of nuclear matter properties with 
observables, there possibly are other correlations
that are intrinsic to nuclear models and the protocols used
to adjust them. Some of the latter correlations might be spurious
consequences of limitations of the models or of the lack of data
that allow to isolate the role of each of the properties of nuclear matter.
For example, it was pointed out early on that the values of $L$ and
$J$ of nuclear EDFs are closely correlated~\cite{(Far78)}, 
which is also found here. Similar correlations are
also found between other elements of the symmetry energy 
\cite{Nazarewicz14a,Mondal17a,Mondal18a}, but their analysis is 
usually limited to bulk properties of infinite matter. It 
has also been pointed out that the volume and surface symmetry energy 
are correlated by nuclear masses through Eq.~\eqref{eq:mac}, see 
for example Refs.~\cite{Nikolov11a,Lattimer13a}. Nuclear masses 
also correlate the surface symmetry energy with 
$\varepsilon_{\text{sat}}$, such that their sum is nearly constant
along the valley of stability~\cite{Nikolov11a}. 
Unfortunately, the portion of the nuclear chart explored experimentally so far
is too small to fix the symmetry parameters in a pure LDM 
model~\cite{(Dob14)}.

It has also been argued that finite nuclei actually 
mainly constrain the symmetry energy $J_{\rho_0}=S(\rho_0)$ and its slope
$L_{\rho_0}$ at sub-saturation densities around 
$\rho_0 \simeq 0.1~\text{fm}^{-3}$~\cite{Ducoin11a,Khan12a,Zhang13a,Horowitz14a}. 
And indeed, as indicated by Fig.~\ref{fig:asurf:JL}, for our nine 
series of fits the values of $J_{0.1}$ and $L_{0.1}$ are somewhat
closer one to each other than those of the corresponding quantity at $\rho_{{\text{sat}}}$.
The spread of these values, however, remains larger than what is typically
found for parametrizations whose $a_{\text{surf}}$ is not constrained, see
for example Ref.~\cite{Ducoin11a} and Table~\ref{tab:INM:ANM} in what follows.

\begin{figure}[t!]
        \includegraphics{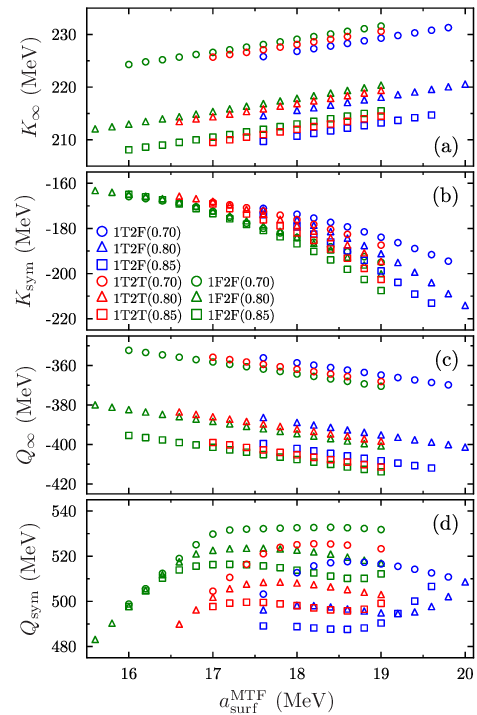}
        \caption{\label{fig:asurf:KQ}
        Higher-order derivatives 
        $K_{\infty}$, $Q_{\infty}$, $K_{\text{sym}}$, and $Q_{\text{sym}}$
        of the binding energy per particle
        at saturation density $\rho_{{\text{sat}}}$.
        Colors and symbols as on Fig.~\ref{fig:asurf:ebya}.
        }
\end{figure}

Figure~\ref{fig:asurf:KQ} displays four higher-order characteristics of infinite matter, which 
are its incompressibility
\begin{equation}
    K_{\infty} = \left.9 \rho_0^2 \, \frac{\partial^2}{\partial \rho_0^2} \,\frac{\mathcal{E}}{\rho_0}\right|_{\rho_0=\rho_\mathrm{sat}} \, ,
\end{equation}
and skewness
\begin{equation}
    Q_{\infty} =\left. 27 \rho_0^3 \, \frac{\partial^3}{\partial \rho_0^3}\, \frac{\mathcal{E}}{\rho_0} \right|_{\rho_0=\rho_\mathrm{sat}} \, ,
\end{equation}
at saturation, as well as the curvature
\begin{equation}
    K_{\text{sym}} = \left.9 \rho_0^2 \, \frac{\partial^2 S(\rho_0)}{\partial \rho_0^2}\right|_{\rho_0=\rho_\mathrm{sat}}
\end{equation}
and skewness
\begin{equation}
    Q_{\text{sym}}=\left. 27 \rho_0^3 \, \frac{\partial^3 S(\rho_0)}{\partial \rho_0^3}\right|_{\rho_0=\rho_\mathrm{sat}}
\end{equation}
of the symmetry energy.
Together with the already discussed coefficients $\varepsilon_{{\text{sat}}}$, 
$J$, and $L$, these parametrize the density dependence of the energy per 
particle and the symmetry energy of symmetric matter around the saturation 
point in terms of $x \equiv (\rho_0-\rho_{{\text{sat}}})/3\rho_{{\text{sat}}}$
\cite{Piekarewicz09,Jerome18_meta_I,Jerome18_meta_II,BaoAnLi_universe21,Grams22}
\begin{align}
\frac{\mathcal{E}(\rho_0)}{\rho_0}
& \simeq \varepsilon_{{\text{sat}}}
        + \tfrac{1}{2} \, K_{\infty} \, x^2
        + \tfrac{1}{6} \, Q_{\infty} \, x^3 
        + \ldots ,
   \\ 
S(\rho_0) 
 & \simeq J 
        + L \, x 
        + \tfrac{1}{2} \, K_{\text{sym}} \, x^2
        + \tfrac{1}{6} \, Q_{\text{sym}} \, x^3 
        + \ldots \, .
\end{align}
The incompressibility $K_\infty$ exhibit a weak linear dependence 
on  $a_{\text{surf}}$ that  is almost independent on the scheme for c.m.\ correction, but 
falls on a different line for each of the three effective masses. The latter finding is a 
consequence of the correlation between $m^*_0/m$, $K_\infty$ and the power $\alpha$ 
of the density-dependent term in the time-even part of the Skyrme EDF of Eq.~\eqref{eq:E:sk:e} that has been identified in 
Ref.~\cite{CHABANAT1997710} and already mentioned in Sec.~\ref{Sect:penalty}. 
The skewness $Q_\infty$ exhibits a similar weak linear dependence 
on  $a_{\text{surf}}$ but in the opposite direction. The reason is that for 
NLO Skyrme EDFs, the equation of state of symmetric matter is entirely determined 
by just three combinations of coupling constants plus the exponent $\alpha$ of the 
density dependence~\cite{CHABANAT1997710}, such that for fixed $\alpha$ and $\rho_\mathrm{sat}$ 
there are only two further linearly independent properties of INM, implying that, 
at given $\varepsilon_{{\text{sat}}}$ and $K_\infty$, the value of $Q_\infty$ is
completely fixed.

The values of $K_{\text{sym}}$ and $Q_{\text{sym}}$, neither of which is
directly constrained in the parameter fit, also change 
over a wide range. This reflects an overall correlation between the density-dependence
of the symmetry energy and the surface energy where changes in one of the symmetry energy's characteristics 
is partially absorbed by changes of the others. For these higher-order coefficients, 
however, the correlation is no longer near-linear over the entire range of values 
for $a_{\text{surf}}$. The values for $K_{\text{sym}}$ show a mild dependence on the scheme for c.m.\ correction only at large $a_{\text{surf}}$, and which is quite different 
from the large offsets found for $J$ and $L$ on Fig.~\ref{fig:asurf:JL}
over the entire range of $a_{\text{surf}}$. The higher-order coefficient 
$Q_{\text{sym}}$, however, exhibits again quite large a dependence on the adopted scheme 
for c.m.\ correction. Both $K_{\text{sym}}$ and $Q_{\text{sym}}$ 
additionally exhibit a mild dependence on $m^*_0/m$, similarly to $J$ and $L$.

\begin{figure}[t!]
        \includegraphics{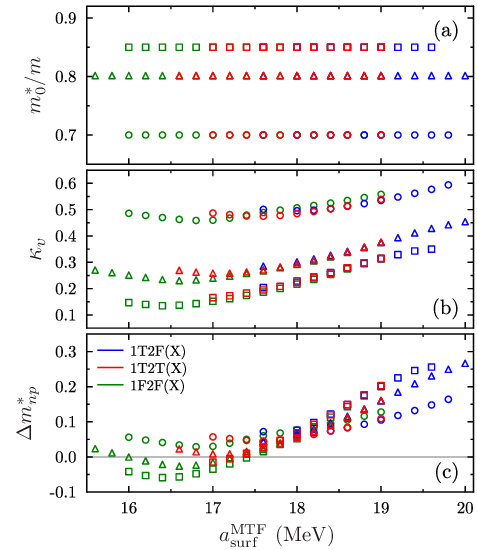}
        \caption{\label{fig:asurf:mstar}
        Isoscalar effective mass $m^*_0/m$, the 
        enhancement factor of the Thomas-Reiche-Kuhn sum 
        rule $\kappa_{v}$, and the splitting $\Delta m^*_{np}$
        of the effective masses of neutrons and protons in neutron matter.
        Colors and symbols as on Fig.~\ref{fig:asurf:ebya}.
        }
\end{figure}

Figure~\ref{fig:asurf:mstar} displays the isoscalar effective mass 
$m^*_0/m$, the enhancement factor $\kappa_{v}$ of the Thomas-Reiche-Kuhn 
sum rule, and the splitting $\Delta m^*_{np} = m^*_n(I)/m - m^*_p(I)/m$ of the effective 
masses of neutrons and protons in pure neutron matter ($I=1$). For 
Skyrme NLO EDFs, these three quantities are not linearly independent, as 
they only depend on two coupling constants of the time-even part of the 
Skyrme EDF of Eq.~\eqref{eq:E:sk:e}
\begin{align}
\frac{m}{m^*_0}
& = 1 + \frac{2m}{\hbar^2} \, C^{\rho \tau}_0 \, \rho_{{\text{sat}}} 
  = 1 + \kappa_s \, , 
    \\
\kappa_v
& = \frac{2m}{\hbar^2} \, \big( C^{\rho \tau}_0 - C^{\rho \tau}_1 \big) 
    \, \rho_{{\text{sat}}} \, ,
    \\
\Delta m^*_{np}
& = \frac{2 (\kappa_v - \kappa_s)}
         {(1 + \kappa_s)^2 - (\kappa_v - \kappa_s)^2}
         \, ,
\end{align}
see Ref.~\cite{Les06} for a detailed discussion. In particular, the 
sign of $\Delta m^*_{np}$ is determined by the sign of $C^{\rho \tau}_1$.

As the isoscalar effective mass is imposed exactly on the respective set of
fits, the panel displaying $m^*_0/m$ mainly serves as a reminder of the colors 
and symbols used to represent the various series of parameter sets.

Contrary to the majority of INM properties discussed so far, for
the new parametrizations constructed here the values for $\kappa_{v}$ and 
$\Delta m^*_{np}$ are strongly correlated to $m^*_0/m$ and 
$a_{\text{surf}}^{\text{MTF}}$, but remain fairly independent on the scheme of 
c.m.\ correction.

At large values of $a_\text{surf}^{\text{MTF}}$, $\Delta m^*_{np}$ 
takes comparatively large positive values and then becomes smaller 
with decreasing $a_\text{surf}^{\text{MTF}}$. For parameter sets 
with $m^*_0/m = 0.7$ the value of $\Delta m^*_{np}$ remains positive 
at all $a_\text{surf}^{\text{MTF}}$, whereas for $m^*_0/m = 0.8$ and 
$m^*_0/m = 0.85$ the values of $\Delta m^*_{np}$ become slightly negative
for the smallest values of $a_\text{surf}^{\text{MTF}}$ covered by our fits.

This property has been analyzed in the context of standard Skyrme NLO EDFs 
before in Ref.~\cite{Les06}. There, it has been pointed out that the 
early Lyon fits such as SLy4-SLy7 and many others yield negative values 
for $\Delta m^*_{np}$, which is at variance with Brueckner-Hartree-Fock (BHF) predictions 
for $\Delta m^*_{np}$ being positive (typically between 0.15 and 0.2 in
pure neutron matter calculated with different flavors of BHF
approximation~\cite{PhysRevC.60.024605}). As argued in Ref.~\cite{Les06}, 
this finding is ultimately caused by the stringent constraints
on the equation of state of neutron matter imposed 
in these fits in combination with a lack of flexibility of the functional 
form of the EDF. For standard Skyrme EDFs with a small power of the density 
dependence $\alpha$ in Eq.~\eqref{eq:E:sk:e}, the contribution of the effective 
mass terms to the total binding energy is strongly constrained by the
high-density regime of the neutron matter equation of state, even if the 
latter's behavior might have a different physical origin. This comes
at the expense of losing the possibility to fine-tune the actual isospin dependence 
of the effective mass, i.e.\ the spectral properties of the single-particle 
Hamiltonian in infinite matter. In fact, the
attempt to push $\Delta m^*_{np}$ to positive values may
even generate parameter sets with finite-size instabilities in the isovector channel.
Adding a second density dependence with sufficiently 
large exponent is one way to resolve these issues~\cite{Les06}. 

We employ here the traditional standard Skyrme EDF with a single density 
dependence that was found to be overconstrained in Ref.~\cite{Les06}.
To have the possibility to freely adjust $a_\text{surf}^{\text{MTF}}$, 
however, we had to substantially relax the constraints on properties of 
infinite nuclear and neutron matter.
As a byproduct, this yields values for the $\Delta m^*_{np}$ that 
are closer to the Brueckner-HF prediction.

\begin{table*}[t!]
\caption{\label{tab:INM:ANM} 
Properties of infinite homogeneous nuclear matter (see text for their definition) of 
the nine best-fit parametrizations. Values for SLy7 and SLy5s1 are shown for 
comparison.
}
\begin{tabular}{lccccccccccccc}
\hline\noalign{\smallskip}
& $\rho_{{\text{sat}}}$ & $\varepsilon_{\text{sat}}$ & $K_{\infty}$  & $Q_{\infty}$ 
& $m^*_0/m$ & $\kappa_v$ & $\Delta m^*_{np}$ &
$J$ & $J_{0.1}$ & $L$ & $L_{0.1}$ & $K_{\text{sym}}$  & $Q_{\text{sym}}$  \\
           & (fm$^{-3}$) & (MeV) & (MeV) & (MeV) & & & & (MeV) & (MeV) &  (MeV)
           &  (MeV) &  (MeV) &  (MeV) \\
\noalign{\smallskip}\hline\noalign{\smallskip}
1T2F(0.70) & 0.160 & $-15.948$ & 229.0 & $-364.1$ & 0.70 & 0.53 &  $+0.10$ & 30.73 & 24.92 & 33.40 & 38.53 & $-182.6$ & 517.0 \\
1T2F(0.80) & 0.160 & $-15.917$ & 216.9 & $-393.0$ & 0.80 & 0.34 &  $+0.11$ & 30.88 & 24.90 & 34.78 & 39.39 & $-184.5$ & 496.3 \\
1T2F(0.85) & 0.160 & $-15.900$ & 211.8 & $-404.7$ & 0.85 & 0.26 &  $+0.13$ & 30.90 & 24.87 & 35.00 & 39.64 & $-186.7$ & 487.6 \\
\noalign{\smallskip}
1T2T(0.70) & 0.160 & $-15.860$ & 228.1 & $-361.9$ & 0.70 & 0.48 &  $+0.05$ & 30.61 & 24.83 & 33.50 & 38.14 & $-176.1$ & 525.1 \\
1T2T(0.80) & 0.160 & $-15.832$ & 216.0 & $-390.7$ & 0.80 & 0.27 &  $+0.03$ & 30.80 & 24.83 & 35.15 & 38.99 & $-175.3$ & 508.2 \\
1T2T(0.85) & 0.160 & $-15.820$ & 211.0 & $-402.6$ & 0.85 & 0.19 &  $+0.03$ & 30.87 & 24.82 & 35.78 & 39.37 & $-175.9$ & 499.5 \\
\noalign{\smallskip}
1F2F(0.70) & 0.160 & $-15.742$ & 226.9 & $-358.7$ & 0.70 & 0.46 & $+0.03$ & 30.56 & 24.82 & 33.35 & 37.90 & $-173.7$ & 531.0  \\ 
1F2F(0.80) & 0.160 & $-15.713$ & 214.7 & $-387.6$ & 0.80 & 0.23 & $-0.03$ & 30.76 & 24.84 & 35.00 & 38.62 & $-170.6$ & 520.6 \\ 
1F2F(0.85) & 0.160 & $-15.701$ & 209.7 & $-399.5$ & 0.85 & 0.14 & $-0.05$ & 30.84 & 24.85 & 35.68 & 38.96 & $-170.0$ & 514.7 \\ 
\noalign{\smallskip}
SLy7       & 0.158 & $-15.894$ & 229.7 & $-358.9$ & 0.69 & 0.25 & $-0.20$ & 31.99 & 25.17 & 47.22 & 41.95 & $-113.3$ & 515.2 \\
SLy5s1     & 0.160 & $-15.772$ & 222.1 & $-372.1$ & 0.74 & 0.30 & $-0.05$ & 31.43 & 24.28 & 48.13 & 42.76 & $-124.8$ & 440.4 \\
\noalign{\smallskip}\hline
\end{tabular}
\end{table*}

The strong correlations between the nuclear matter properties examined above 
can have two clearly distinct reasons. On the one hand, observables of finite nuclei 
are known to be only sensitive to specific combinations of two or more properties of INM.
On the other hand, the number of INM properties we analyze here is actually larger than the
number of parameters of the Skyrme EDF at NLO that determine them.
In particular, $m^*_0/m$, $\kappa_v$
and $\Delta m^*_{np}$ only depend on parameters of the terms with gradients in the Skyrme
generator that also make a large contribution to $a_\text{surf}$~\cite{Ryssens19a}
and $a_\text{ssym}$. The space for these parameters however is limited by the appearance of
finite-size instabilities in the four $(S,T)$ channels.

These observations indicate that over-constraining 
some specific properties of nuclear bulk matter in the parameter 
adjustment of an EDF with a limited number of degrees of freedom 
can lead to very unrealistic results for other properties. The latter can
be higher-order characteristics of homogeneous matter, features of 
inhomogeneous matter in general, or more specifically surface properties.
Indeed, it has been pointed out before that the parametrizations of the 
standard Skyrme EDF that reproduce best the knowledge about nuclear matter
properties of the time~\cite{(Dut12a)} do not well describe finite nuclei 
\cite{Stevenson13a}; surface properties are probably only one aspect
of this puzzle. Conversely, extended Skyrme EDFs are needed to 
describe the global systematics of nuclear masses and the
present empirical knowledge about neutron stars within a single model
\cite{Goriely10a,Goriely13c,grams23a}.
This observation suggests that constraining nuclear matter properties 
at densities and asymmetries that are far from those encountered in 
finite nuclei does not necessarily fix loose ends in the parametrization of 
a given EDF tailored for the description of finite nuclei as sometimes 
hoped for, but leads to independent properties
that cannot be simultaneously modeled within the same simple form of the EDF.
These concerns can all be traced to the limited number of degrees of freedom of the 
standard Skyrme EDF. Reconciling some or all of these issues will require 
extending the form of the EDF, whether through additional density 
dependencies~\cite{Cochet04a,Les06}, combined momentum and density dependencies
\cite{Krewald77,Goriely10a,Goriely13c,Zhen16a,grams23a},
or higher-order momentum dependent terms \cite{Raimondi11a,Becker17}.

%
%
\subsection{Fits without constraint on $a_\text{surf}^\text{MTF}$}
\label{sect:penalty:best}

Since the optimization of the parameters for the EDFs of type 1T2T
gives the lowest $\chi^2$ for a value of $a_\text{surf}^\text{MTF}$ 
that is close to the expected optimal value 
to describe the properties of nuclei at large deformation, we 
added one additional parametrization to each series of each type without 
a constraint on $a_\text{surf}^\text{MTF}$. We call these parametrizations 
``best fits'' in what follows, but underline that they only represent a best 
fit with respect to the penalty function defined in Sec.~\ref{sect:adjustment}
for a given choice of c.m.\ correction and $m^*_0/m$; these fits are not
necessarily optimal to describe nuclear deformation properties.
The coupling constants of these parametrizations 
can be found in the supplementary material~\cite{supplement}.
As the surface properties of these additional fits are not 
constrained by information on deformed nuclei, they can also be used for
a study of the impact of the choice for c.m.\ correction and effective mass
on deformation properties of existing parametrizations.

\begin{table}[t!]
\caption{\label{tab:SINM} 
Properties of semi-infinite nuclear matter as obtained with the nine ``best-fit'' 
parametrizations (all in units of MeV), where the last two columns list the
effective surface energy coefficients $a^{\text{HF}}_{\text{surf,eff}}(I)$ obtained
in HF approximation at the asymmetries of the \nuc{180}{Hg} ($I=0.111$)
and \nuc{240}{Pu} ($I=0.217$) nuclei discussed in Sec.~\ref{sect:deformation}.
Values for SLy7 and SLy5s1 are shown for comparison.
}
\begin{tabular}{lcccccc}
\hline\noalign{\smallskip}
& & & & & 
    \multicolumn{2}{c}{$a^{\text{HF}}_{\text{surf,eff}}$}  \\
    \noalign{\smallskip}\cline{6-7}\noalign{\smallskip}
&  $a_{\text{surf}}^{\text{MTF}}$ &  $a_{\text{surf}}^{\text{ETF}}$ &  $a_{\text{surf}}^{\text{HF}}$ &
    $a_{\text{ssym}}^{\text{HF}}$ 
& \nuc{240}{Pu} & \nuc{180}{Hg} \\
\noalign{\smallskip}\hline\noalign{\smallskip}
1T2F(0.70) & 18.9 & 18.0 & 18.4 & $-49$ & 16.1 & 17.8 \\
1T2F(0.80) & 18.6 & 17.8 & 18.2 & $-51$ & 15.8 & 17.6 \\
1T2F(0.85) & 18.4 & 17.7 & 18.1 & $-51$ & 15.7 & 17.5 \\
\noalign{\smallskip}
1T2T(0.70) & 18.0 & 17.1 & 17.5 & $-47$ & 15.3 & 16.9 \\
1T2T(0.80) & 17.7 & 16.9 & 17.3 & $-49$ & 15.0 & 16.7 \\
1T2T(0.85) & 17.6 & 16.9 & 17.2 & $-50$ & 14.9 & 16.6 \\
\noalign{\smallskip}
1F2F(0.70) & 17.1 & 16.2 & 16.6 & $-44$ & 14.5 & 16.1 \\ 
1F2F(0.80) & 16.8 & 16.0 & 16.4 & $-46$ & 14.3 & 15.9 \\ 
1F2F(0.85) & 16.7 & 15.9 & 16.3 & $-47$ & 14.1 & 15.8 \\ 
\noalign{\smallskip}
SLy7       & 18.0 & 17.1 & 17.5 & $-51$ & 15.1 & 16.9 \\
SLy5s1     & 18.0 & 17.1 & 17.6 & $-56$ & 14.9 & 16.9 \\
\noalign{\smallskip}\hline
\end{tabular}
\end{table}

The nuclear matter properties of these parametrizations are listed 
in Table~\ref{tab:INM:ANM}. As can be expected from the previous 
discussion of the correlations between nuclear matter properties and 
$a_{\text{surf}}$, and from the systematic differences between the values 
of $a_{\text{surf}}$ at the minimum of the penalty function shown in 
Fig.~\ref{chi2}, the nuclear matter properties of the nine ``best fits''
vary over a wide range  of values, including those constrained in the 
fit. The values for $\varepsilon_{\text{sat}}$ fall inside the tolerance interval 
of the penalty function only for the 1T2F(X), and the values for $J$ and $L$
even systematically fall outside the tolerance interval of the penalty 
function for all of them.

Table~\ref{tab:INM:ANM} also lists the symmetry energy 
$J_{0.1}$ and its slope $L_{0.1}$ at $\rho_{0} = 0.1~\text{fm}^{-3}$.
These two quantities at this sub-saturation density are more stringently constrained by the 
properties of finite nuclei than the values of $J$ and $L$ at saturation 
density, as we mentioned before in the context of Fig.~\ref{fig:asurf:JL}.
The same is found for the nine ``best fits''. In particular, the 
$J_{0.1}$ values of the 1F2F(X) and 1T2T(X) fits are near-identical within 
a few tens of keV, although their values for $J$ differ by several hundreds 
of keV. The values of the symmetry energies $J_{0.1}$ of the three 1T2F(X) 
fits are also much closer than their $J$, but remain slightly larger 
than those of the 1F2F(X) and 1T2T(X) fits.

As expected from Fig.~\ref{fig:asurf:mstar}, all 1T2F(X) and 1T2T(X) parameter 
sets predict a positive splitting of $\Delta m^*_{np}$, in agreement with 
Brueckner-HF calculations. Only the 1F2F(X) take negative values as did
the earlier SLyX parameter sets~\cite{Les06}.

Table~\ref{tab:INM:ANM} also lists the nuclear matter properties
of the SLy7 \cite{CHABANAT1998231} and SLy5s1 \cite{Jodon16a} parametrizations 
that were adjusted with similar, albeit not identical, protocols as ours, and that
are both known to provide a reasonable description of fission barriers. There
are some noteworthy differences in their nuclear matter properties. The SLy7 parameter
set is of 1T2T type and can be directly compared with 1T2T(0.70) that has almost the same 
effective mass. For SLy7, the values of $J$ and $L$ are actually closer to the target 
values of our fit protocol then they are for any of our best fits. This, however, 
comes at the expense of SLy7 exhibiting finite-size spin instabilities, such that it
cannot be used in calculations that break time-reversal symmetry without making an 
ad hoc modification of the $C^{s \Delta s}_t$ coupling constants in the time-odd 
part~\eqref{eq:E:sk:o} of the Skyrme EDF. 
This is different for SLy5s1 that has already been adjusted with a constraint
on the absence of unphysical finite-size instabilities at densities probed in finite 
nuclei. The SLy5s1 parameter set is of 1T2F type and has an effective mass that 
falls in between the values of 1T2F(0.70) and 1T2F(0.80). Again, SLy5s1 yields 
values for $J$ and $L$ that are also closer to the targeted values of our fit protocol.
However, its value for $\varepsilon_{\text{sat}}$ is much further away. These 
differences result from slight differences in the choices made when setting up the penalty
function and indicate that for overall well-adjusted parametrizations of the 
standard Skyrme EDF any significant improvement with respect to one nuclear matter 
property can in general only be achieved when significantly degrading others. 

\begin{figure}[t!]
\includegraphics{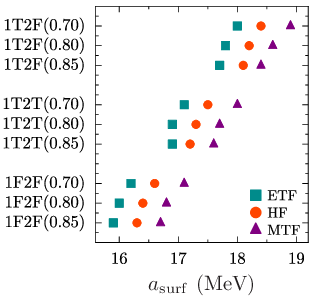}
\caption{\label{asurf}
Value of the surface energy coefficient $a_{\text{surf}}$ 
calculated with the HF, ETF and MTF methods for the nine
``best fits''.
}
\end{figure}

Table~\ref{tab:SINM} lists surface properties of the nine ``best fits'' as
obtained from calculations of semi-infinite matter. The first three columns
provide the surface energy coefficient $a_{\text{surf}}$ calculated with the 
HF, ETF and MTF methods, which are also illustrated on Fig.~\ref{asurf}. 
There is again a near-constant shift between the methods with a slight
effective-mass dependence, as could be expected from the analysis of 
the parametrizations with systematically varied 
$a_{\text{surf}}^{\text{MTF}}$ as shown on Fig.~\ref{fig:asurf:correlation}.

As the parameters of the nine ``best fits'' correspond to the minima of the penalty 
functions plotted on Fig.~\ref{chi2}, the value of
their surface energy coefficient depends strongly on their respective 
scheme for c.m.\ correction and also their isoscalar effective mass $m^*_0/m$.
Choosing a different scheme for c.m.\ correction leads to significantly 
different values of $a_{\text{surf}}$. Confirming the earlier analysis 
of Ref.~\cite{(Ben00f)}, those of the 1T2F(X) are typically almost 1~MeV 
larger than those of the 1T2T(X), whereas those for the 1F2F(X) are 
about 1~MeV smaller than those for the 1T2T(X), which will make an 
enormous difference for fission barriers.

To a lesser extent, choosing a different effective mass also 
yields significantly different values for $a_{\text{surf}}$ when 
not constraining the latter. Within a series of fits with same scheme for 
c.m.\ correction and compared with the fit with $m^*_0/m = 0.7$, the 
value of $a_{\text{surf}}$ of the fit with $m^*_0/m = 0.8$ is 
about 200~keV smaller, and the one of the fit with $m^*_0/m = 0.85$
even about 300~keV smaller. 
As discussed in Refs.~\cite{Jodon16a,Ryssens19a}, changing 
$a_{\text{surf}}$ by as little as 0.2~MeV typically changes the 
outer fission barrier of \nuc{240}{Pu} on the order of 700~keV.

Table~\ref{tab:SINM} also lists the surface-symmetry energy coefficient 
$a_{\text{ssym}}^{\text{HF}}$ calculated in HF approximation as well 
as the effective surface symmetry coefficient $a_{\text{surf,eff}}(I)$ from Eq.~(\ref{eq:asurfeff})
of the two nuclei \nuc{240}{Pu} (with $I=0.217$) and 
\nuc{180}{Hg} (with $I=0.111$), whose fission barrier properties will be analyzed in 
Sec.~\ref{sect:deformation}.

For the rest of the discussion, we will focus on these parametrizations,
labeled 1F2F(X), 1T2F(X), 1T2T(X), and that can be expected to be 
representative for the typical behavior of standard Skyrme interactions 
adjusted with a given scheme for c.m.\ correction at a given effective mass.

%
%
\subsection{The origin of the correlations between nuclear matter properties}

\begin{figure}[t!]
    \includegraphics{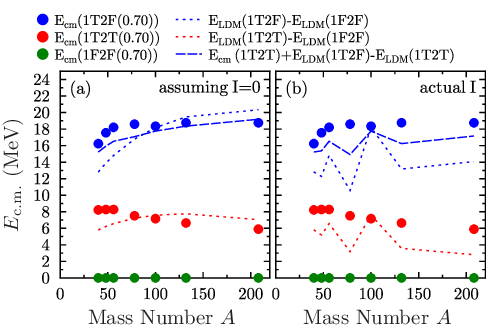}
\caption{\label{origin} 
Size of the c.m.\ correction energy $E_{\text{c.m.}}$ (full markers) for the 
three parametrizations with $m^*_0/m = 0.70$ as indicated for the seven 
doubly-magic nuclei entering the penalty function of the new fits 
plotted as a function of their mass number. The lines are estimates
for the size of the c.m.\ correction based on differences of the LDM 
coefficients obtained for the three fits  (see text). 
Panel (a) compares with the LDM estimates assuming $I=0$ for all nuclei, i.e.\ 
considering only the volume and surface energy, whereas for the lines
in panel (b) also the symmetry and surface symmetry contributions
to the LDM energies are taken into account.}
\end{figure}

As already mentioned, it has been pointed out before~\cite{(Ben00f)} that the
significantly different values of $a_{\text{surf}}$ obtained in fits that (i) use different
schemes for c.m.\ correction and that (ii) are only constrained by data on spherical nuclei or
nuclear matter, results from the nuclear matter properties absorbing the absent contribution
from the c.m.\ correction energy to the total binding energy of the nuclei entering 
the penalty function during the parameters adjustment.

For the seven doubly-magic nuclei entering the adjustment protocol,
the size of the c.m.\ correction energy $E_{\text{c.m.}}$ is displayed in 
Fig.~\ref{origin} for the three parametrizations with $m^*_0/m = 0.70$ by
filled markers.
The full c.m.\ correction of the 1T2T(0.70) parametrizations only takes about one 
third of the size of the one-body contribution of the 1T2F(0.70). The 
$A$-dependence of the c.m.\ correction energy is also different in the three 
cases: for the 1F2F(0.70) it is constant and zero by construction, for the 
1T2F(0.70) it quickly rises for light nuclei and then remains almost 
constant for the heavy ones,\footnote{Note that the one-body contribution
$E_{\text{c.m.}}^{(1)}$ to the c.m.\ correction energy does not fall off to zero
in the limit $A \to \infty$; only the sum of the one-body and two-body 
contributions does for reasons evoked in the introduction. 
Instead, when increasing $A$ beyond the interval shown on Fig.~\ref{origin}, 
the value of $E_{\text{c.m.}}^{(1)}$ tends to a value that equals the contribution 
of the kinetic energy to the \textit{energy per particle} in infinite matter, 
which for symmetric matter is
$\tfrac{3}{5} \, \tfrac{\hbar^2}{2m} \, k_{\text{F}}^2
= \tfrac{3}{5} \, \tfrac{\hbar^2}{2m} \big( \tfrac{3\pi^2}{2}  \rho_{\text{sat}} \big)^{2/3}
= 22.108 \, \text{MeV}$.
} 
whereas for the 1T2T(X) it 
slowly falls off with mass number (when plotting $E_{\text{c.m.}}$ 
for all nuclei across the chart one also clearly sees shell effects 
introduced by the two-body contribution~\cite{(Ben00f)}).
Figure~\ref{origin} clearly indicates that neglecting the two-body 
contributions to the c.m.\ correction for reasons of computational 
convenience neither constitutes a quantitatively nor a qualitatively 
meaningful approximation. It is because of the different non-linear 
$A$ dependence of the resulting c.m.\ correction energy that different 
schemes for the c.m.\ correction have a large impact on the surface 
energy when fitting parameter sets.

This is illustrated by the three lines on Fig.~\ref{origin}.
We recall that we use a convention~\eqref{eq:corr} for $E_{\text{c.m.}}$ where 
it enters the total binding energy \eqref{eq:Etot} with a minus sign.
Assuming that the LDM formula perfectly simulates the
contributions from the kinetic, Skyrme and Coulomb energies to the total
energy of nuclei when inserting the nuclear matter properties of a given
parametrization, one should find 
$E_{\text{LDM}}^{\text{1F2F}} = E_{\text{LDM}}^{\text{1T2T}} - E_{\text{c.m.}}^{\text{1T2T}}$ or, equivalently,
\begin{equation}
E_{\text{c.m.}}^{\text{1T2T}} 
= E_{\text{LDM}}^{\text{1T2T}} - E_{\text{LDM}}^{\text{1F2F}} 
\end{equation}
and similar when comparing two other pairs of parametrizations. The dotted lines
on Fig.~\ref{origin}
plot the difference between the LDM energies, calculated through Eq.~\eqref{eq:mac} 
and using the nuclear matter properties listed in Tables~\ref{tab:INM:ANM} 
and~\ref{tab:SINM}, of either the 1T2T(0.70) and 1F2F(0.70) (drawn in red) or 
the 1T2F(0.70) and 1F2F(0.70) (drawn in blue) parametrizations, respectively. 
In panel~(a) only the isoscalar volume and surface terms of the LDM energy
are included in this analysis, whereas in panel~(b) this is done for the full 
LDM energy including the symmetry and surface symmetry terms.\footnote{Note that the 
Coulomb energy does not contribute to these LDM estimates as, by construction,
$\rho_{\text{sat}}$ is the same for all parametrizations considered here.} 
The dashed blue line shows the sum of the c.m.\ correction energy 
$E_{\text{c.m.}}$ obtained with 1T2T(0.70) and the difference between the 
LDM  energies obtained with 1T2F(0.70) and 1T2T(0.70). If the contribution
of $E_{\text{c.m.}}$ to the total binding energy was perfectly absorbed by 
the nuclear matter properties, then the three lines would fall on top of
the markers of same color.

For the simpler estimate made in panel~(a) this is almost the case; in particular 
the difference in c.m.\ correction energy between 1T2T(0.70) and 1T2F(0.70) is very 
well reproduced by the LDM estimate, as already observed in Ref.~\cite{(Ben00f)}. 
The absolute size of the c.m.\ correction energies found with 1T2T(0.70) and 
1T2F(0.70), however, is less well described by the difference in LDM energies 
between either and 1F2F(0.70). Also, including the isovector terms in the LDM energy, 
in particular the surface symmetry energy whose coefficient varies by several MeV, 
somewhat spoils the agreement between $E_{\text{c.m.}}$ and the LDM estimates, 
pointing to a more complex compensation between terms in the parameter adjustment 
as far as the isovector degree of freedom is concerned. This is not too surprising 
as the LDM expression for the energy assumes that the isovector density is constant 
throughout the nucleus, and the symmetry energy the same at all densities, 
which is not at all the case in a self-consistent mean-field model.

Following Ref.~\cite{(Ben00f)}, the overall size and sign of the differences 
between the nuclear matter properties of the 1F2F(X), 1T2F(X), and 1T2T(X) 
can be explained by fitting a simplified LDM expression 
$b_{\text{vol}} \, A + b_{\text{surf}} \, A^{2/3}$ directly to the 
c.m.\ correction energies plotted on Fig.~\ref{origin}. For 1T2T(0.70) one finds
$b_{\text{vol}} = -0.185$ MeV, $b_{\text{surf}} = 1.241$ MeV.
These numbers are of similar size as the differences 
$\Delta a_{\text{vol}} = -0.118$ MeV and $\Delta a_{\text{surf}} = 0.9$ MeV
found between the values of these coefficients for 1T2T(0.70) and 1F2F(0.70)
in Tables~\ref{tab:INM:ANM} and~\ref{tab:SINM}. A similar qualitative agreement 
is found for 1T2F(0.70) and 1F2F(0.70) with $b_{\text{vol}} = -0.325$ MeV, $b_{\text{surf}} = 2.431$ MeV and
$\Delta a_{\text{vol}} = -0.206$ MeV, $\Delta a_{\text{surf}} = 1.8$ MeV. 

That $a_{\text{vol}}$ and $a_{\text{surf}}$ have to change simultaneously
in opposite direction when the interaction energy has to absorb the contribution
from the c.m.\ correction to the binding energy becomes evident when considering 
the c.m.\ correction energy to be roughly independent of $A$ for the nuclei entering the 
adjustment protocol. A constant change in binding energy of these nuclei 
can be roughly achieved by a small change of the volume term $\propto A$ and 
a larger change of the surface term $\propto A^{2/3}$ in the opposite
direction. For example, assuming that $E_{\text{c.m.}}$ of the 1T2T(0.70) 
parametrization is simply 7~MeV for nuclei in the range $40 \leq A \leq 208$, 
a least-square fit of the simplified LDM formula to these values yields 
$b_{\text{vol}} = -0.135$ MeV, $b_{\text{surf}} = 0.99$ MeV, which is even closer
to the actual change of the nuclear matter properties when comparing 1F2F(0.70)
with 1T2T(0.70) than what is found fitting the precise values for $E_{\text{c.m.}}$
obtained for 1T2T(0.70).
Repeating the same estimate with 19~MeV as an approximation for the c.m.\ correction 
energy of the 1T2F(0.70) parametrization leads to $b_{\text{vol}} = -0.348$ MeV and $b_{\text{surf}} = 2.53$ MeV.

As said before, taking into 
account that some of the nuclei are asymmetric leads to a less clear picture.
We also recall that  $a_{\text{vol}} = \varepsilon_{\text{sat}}$ and 
$a_{\text{sym}} = J$ are constrained in our adjustment protocol, such that 
their values cannot vary freely when fitting parametrizations with different
schemes for c.m.\ correction. In particular, the above analysis indicates that 
$a_{\text{vol}} = \varepsilon_{\text{sat}}$ has to change by about the size of 
its tolerance in the adjustment protocol in order to simulate the presence 
or absence of one or the other contribution to the c.m.\ correction.

These findings are consistent with the presumption of Ref.~\cite{(Ben00f)} that the 
absent contributions from the c.m.\ correction to the total binding energy of the 
nuclei considered in the fit are absorbed by the nuclear matter properties of the resulting parameter 
sets, and demonstrates that it also applies to the comparison with parameters sets 
that do not consider any c.m.\ correction at all.

Results found for the parametrizations with $m^*_0/m = 0.80$ and $m^*_0/m = 0.85$ are very
similar to what is shown on Fig.~\ref{origin}, with a subtle difference in detail that would,
however, be difficult to identify on a plot: for the seven nuclei entering the fit, 
the c.m.\ correction energy calculated in either the 1T2T or the 1T2F scheme decreases by 
roughly 200~keV when going from a fit with  $m^*_0/m = 0.70$ to a fit with $m^*_0/m = 0.85$. 

We also mention in passing that, for the 1T2T(X) fits, the one-body contribution 
to the c.m.\ correction is typically 200 keV larger than the value found with the 1T2F(X) 
fit with same effective mass because of self-consistency effects.

Compared with other contributions to the binding energy, a particularity 
of the c.m.\ correction energy is that over the range of experimentally 
accessible nuclei it is almost 
constant. It turns out that modern refined liquid-drop models also contain large 
mass-independent terms, i.e.\ have contributions that are proportional to $A^0$. 
These terms still have an isospin dependence, though. In the finite-range liquid-drop model (FRLDM) of Ref.~\cite{Moller16a} there 
is an explicit $A^0$ term as well as a contribution from the finite-range surface 
energy that scales as $A^0$. In the notation of that paper, for spherical nuclei 
the sum of these terms is given by 
$a_0 - \tfrac{3a^2}{r_0^2} a_s (1 - \kappa_s I^2)$. Inserting the values of the 
constants, one finds $2.645 - 21.927 (1 - 2.39 \, I^2)$~MeV, which varies between 
$-19.28$~MeV for \nuc{40}{Ca} and other $N=Z$ nuclei and $-16.93$~MeV for \nuc{208}{Pb}. 
These values are very close to the contribution of the one-body term of the 
c.m.\ correction to the total energy as obtained with 1T2F(0.70) and 
plotted on Fig.~\ref{origin}. 
The Lublin-Strasbourg liquid drop model (LSD) of Ref.~\cite{Pomorski03a} 
also has a sizable contribution $\propto A^0$, which there is motivated as a Gaussian 
curvature term. In the notation of the paper, it takes the form 
$b_{\text{curG}}(1-\kappa_{\text{curG}}\, I^2) A^0$. Inserting again the values 
of the constants for the NLD parametrization of the LSD, one finds 
$10.357 \, (1 - 13.4235 \, I^2)$, which takes a positive value 
of 10.357 MeV for all $N=Z$ nuclei, and falls off to 4.14~MeV for heavy nuclei on 
the valley of stability like \nuc{208}{Pb}, and even might become slightly 
negative for very neutron-rich ones, such as $-0.703$~MeV for \nuc{78}{Ni}. 
The $A^0$ terms of these modern liquid-drop models therefore behave very 
differently. Although such term-by-term comparisons between different approaches
have to remain qualitative, it seems that the successful reproduction of nuclear masses 
requires the presence of rather large contributions to the binding energy 
that are essentially mass-independent.

%
%
\section{Deformation properties}
\label{sect:deformation}

%
\subsection{Set-up of the calculations}

To explore the deformation properties of the newly constructed parametrizations, 
we turned to the \texttt{MOCCa} code of Ref.~\cite{ryssens_2016}, which represents
single-particle wave functions on a three-dimensional Cartesian 
coordinate-space mesh with equidistant points. Profiting from the efficiency 
of Lagrange meshes~\cite{baye_2015}, the relatively coarse discretisation 
$dx = 1.0$~fm chosen for this study is sufficiently accurate to resolve 
absolute binding energies to within a few hundred keV~\cite{Ryssens15a}.
This is sufficient for our purposes, especially since this numerical precision 
is essentially independent of the nuclear shape, even for the 
very elongated shapes we discuss below~\cite{Ryssens15a}, such that 
we expect differences of binding energies to be even more accurately 
resolved.

The deformation of the nuclear density can be characterized by its multipole 
moments $Q_{\ell m}$. For two integers $\ell$ and $m$ that satisfy
$\ell \geq m \geq 0$, we define
\begin{equation}
\label{eq:qlm}
Q_{\ell m} 
= \int \! \mathrm{d}^3r \, \rho_0(\vec{r}) \, r^{\ell} \, \text{Re} \left[ Y_{\ell m}(\theta, \phi)\right] \, ,
\end{equation}
where $\rho_0(\vec{r})$ is the matter density and $Y_{\ell m}(\theta, \phi)$ is a spherical harmonic.
Since the $Q_{\ell m}$ scale with particle number, it is more straightforward
to compare dimensionless multipole moments $\beta_{\ell m}$:
\begin{equation}
\label{eq:betalm}
\beta_{\ell m } 
= \frac{4 \pi } {  3  R^{\ell}  A } Q_{\ell m} \, , 
\end{equation}
where $R = 1.2 \, A^{1/3}$ fm. By replacing $\rho_{0}(\vec{r})$ in Eq.~\eqref{eq:qlm} 
by the neutron or proton density and replacing $A$ in Eq.~\eqref{eq:betalm}
by either $N$ or $Z$, we also define the neutron and proton multipole moments
$\beta_{q, \ell m}$ with $q=p$, $n$. We will in what follows assume that the 
nuclear charge density equals the proton density, which implies that 
the charge and proton multipole moments are equal.

The flexibility of the \texttt{MOCCa} code with respect to the symmetries
imposed on the eigenstates of the single-particle Hamiltonian
is used to reduce the computational effort. All calculations reported here
conserve time-reversal symmetry, $z$ signature $\hat{R}_z$ and the $y$ 
time-simplex $\hat{S}^T_y$. The combination of the latter two imposes two plane symmetries in the $x=0$ 
and $y=0$ planes on the local densities and currents~\cite{Ryssens15b}. 
For the calculation of fission barriers at large deformation, parity 
$\hat{P}$ is not enforced, which allows for the description of shapes that 
are not reflection symmetric with respect to the $z=0$ plane. In this case,
a constraint on the mass dipole moment $\beta_{10}$ is added to fix the 
nucleus' center-of-mass at the origin of the numerical box.
For the study of shapes at small deformation, however, it turned out that
for the majority of cases parity can be enforced as a conserved symmetry 
without loss of generality. This reduces the computational cost and 
facilitates the convergence of the self-consistent equations. For either of these two choices, the
Cartesian 3d representation allows for the description of non-axial shapes.
It turned out, however, that most of the states discussed below remain axially symmetric.

%
%
\subsection{Treatment of pairing correlations}
\label{sec:mocca_pairing}

The 1F2F(X), 1T2F(X), and 1T2T(X) parametrizations were adjusted to 
properties of doubly-magic nuclei 
for which pairing correlations vanish at the mean-field level. The 
calculations of energy surfaces and deformed open-shell nuclei that
will be presented in what follows, however, require the introduction 
of pairing correlations. These are treated by solving the HFB equations 
within the two-basis method~\cite{(Ga94),(Rys19b)}. 
For the effective pairing EDF, we employ the widely-used density-dependent 
form~\cite{(Rig99)}
\begin{equation}
E_{\mathrm{pair}} 
= \sum_{q =p,n} \frac{V_q}{4} 
  \int \! \mathrm{d}^3r \,
\bigg[1 - \frac{\rho_0(\vec{r})}{\rho_c} \bigg] \, 
\tilde{\rho}_q^{*}(\vec{r}) \, \tilde{\rho}_q(\vec{r}) \, ,
\end{equation}
where $\tilde\rho_q(\vec{r})$ is the pair density~\cite{Bender03r}.
With $\rho_c = 0.16$ fm$^{-3}$, this form corresponds to a ``surface-type''
pairing interaction. A smooth cutoff above and below the Fermi energy 
as described in Refs.~\cite{(Rys19b),Jodon16a,Ryssens19a} limits the pairing
correlations to the single-particle levels around the Fermi energy. For simplicity, 
we take the proton and neutron pairing strengths to be equal, i.e. $V_n = V_p = V_1$, and 
use the same cutoff parameters ($\mu_p = \mu_n = 0.5$~MeV and $\Delta E_p = \Delta E_n = 5.0$~MeV) for both species,
as done before in Refs.~\cite{(Rig99),(Rys19b),Jodon16a,Ryssens19a}.

While our earlier studies on deformation energies reported in 
Refs.~\cite{Jodon16a,Ryssens19a} used the HFB+Lipkin-Nogami (LN) scheme to ensure the
presence of pairing correlations in all states, we use here the
stabilisation of the pairing EDF proposed in Ref.~\cite{Erler08a}
instead, with $E_{\text{cutb}} = 0.3$~MeV for the cutoff parameter. 

It is well-known that the pairing strength has to scale with 
effective mass $m^*_0/m$. For the series with $m^*_0/m = 0.70$
we use the same pairing strength of $V_1 = -1250~\mathrm{MeV}\,\mathrm{fm}^{-3}$
for protons and neutrons originally adjusted for SLy4 in 
Ref.~\cite{(Rig99)} and used in Refs.~\cite{Jodon16a,Ryssens19a}. 
Although originally adjusted within the HFB+LN scheme, this 
pairing strength gives nearly identical values of pairing gaps 
when used in the context of the stabilized pairing EDF.
For the two other series, the pairing strength was readjusted
to give the same average neutron pairing gap for the spherical 
ground state of \nuc{188}{Pb} as SLy4 with $V_1 = -1250~\mathrm{MeV}\,\mathrm{fm}^{-3}$,
which led to the values of $V_1 = -1175~\mathrm{MeV}\,\mathrm{fm}^{-3}$ for the
parametrizations with $m^*_0/m = 0.80$ and $V_1 = -1140~\mathrm{MeV}\,\mathrm{fm}^{-3}$
for those with $m^*_0/m = 0.85$. 

In some figures, we compare results obtained with the new fits with 
results obtained with the existing SLy7~\cite{CHABANAT1998231} and SLy5s1 
\cite{Jodon16a} parametrizations that are known to have reasonable 
deformation properties. Both have an isoscalar effective mass close to 0.7, 
see Table~\ref{tab:INM:ANM}, and will be used with a pairing strength of 
$V_q = -1250~\mathrm{MeV}\,\mathrm{fm}^{-3}$.

%
%
\subsection{Fission barrier of $^{240}$Pu}
\label{sect:Pu240:barrier}

\begin{figure}[t!]
\includegraphics{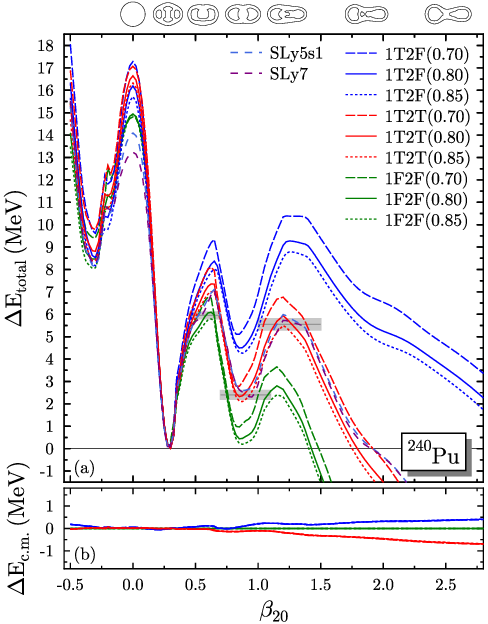}
\caption{\label{Pu240} 
Deformation energy curve of $^{240}$Pu (panel (a)) and change of the
center-of-mass correction (panel (b)) as a function of the quadrupole 
deformation $\beta_{20}$ for the parametrizations as indicated.
In both cases, the energies are normalized to the value at the respective 
ground-state deformation. To facilitate the comparison, both panels share 
the same energy scale.
The horizontal grey bars in panel (a) indicate experimental values for 
the height of the inner and outer barriers as well as the excitation 
energy of the fission isomer, taken from the sources mentioned in the text. The inserts on top of the figure indicate the evolution of 
shapes along the fission path, the upper and lower halves representing isodensities at $\rho_0=0.08$
and 0.15~fm$^{-3}$ in the $x$ and $y$ directions.}
\end{figure}

We start our discussion of deformation properties by considering
the double-humped fission barrier of $^{240}$Pu, which is arguably the most
widely-used testing ground for the modeling of nuclear
fission~\cite{Bender03r,Jodon16a,Bartel82a,Bender04a,Berger89_D1S,PhysRevC.69.014307,Bonneau04,Younes09,Li10,Abusara12,Schunck14,Rutz95,Samyn05}.
Figure~\ref{Pu240} displays the static fission barrier of 
this nucleus calculated as in
Ref.~\cite{Ryssens19a}. For all eleven parametrizations, 
we find a very similar fission path in the space of multipole deformations 
$\beta_{\ell m}$ that evolve continuously without sudden jumps. There is
one little difference in detail for 1F2F(0.70), 1F2F(0.80), 1F2F(0.85), 
1T2T(0.70) and SLy7. For these we find a narrow region around the ground 
state where octupole deformation leads to a small additional energy gain: 
200~keV for 1F2F(0.70) and a few tens of keV for the four others. For
all other parametrizations, all configurations are reflection symmetric up to 
the superdeformed minimum associated with the fission isomer.
At larger quadrupole deformations, octupole deformation gradually sets in
and shapes become reflection-asymmetric. Around the two saddle
points, the lowest-energy path passes through non-axial shapes that lower the 
inner barrier by about 1.5~MeV and the outer one by about 0.5~MeV as 
found earlier in Refs.~\cite{Ryssens19a,Ryssens23iib}. 
The corresponding $\beta_{22}$ deformation takes values of about 0.07  
for the inner and 0.02 for the outer barrier, which corresponds 
to $\gamma$ angles of about 12~degrees and 1.5~degrees, 
respectively. At small deformations and around the minima, the nucleus 
takes an axial shape. Altogether, the fission path is very similar to the one of the actinide nuclei discussed in Ref.~\cite{Ryssens23iib}.

The energy curves obtained with the new parameter sets fall into three clearly
distinguishable groups that are identified by the scheme for c.m.\ correction 
employed during their adjustment: 
those using the 1T2F recipe give systematically the highest energy curves relative to the ground state when increasing deformation,
those using the 1F2F recipe the lowest ones,
while the 1T2T sets 
fall in between. The systematic differences are enormous: compared with 
the 1T2T(X) parameter sets, the excitation energy of the fission isomer is 
about 2~MeV larger for those in the 1T2F(X) set, while for the 1F2F(X) sets 
it is about 2~MeV smaller. 
For the height of the outer barrier, the differences are 
even larger. Within each of these three groups, there also is a clear 
dependence of the deformation energy on isoscalar effective mass: 
for a given recipe of c.m.\ correction, the deformation energy systematically 
increases with decreasing $m^*_0/m$, and this in a very similar way 
for each of the three recipes. 

As can be seen from panel (b) of Fig.~\ref{Pu240}, the variation of the 
c.m.\ correction energy $E_{\text{c.m.}}$ with deformation is much smaller 
than the difference  between the barriers and therefore cannot explain it.
Still, the one-body contribution systematically increases the barriers 
by a few 100~keV, whereas the full c.m.\ correction reduces the barriers by
a few 100~keV. The value of the effective mass has practically no 
influence on the variation of $E_{\text{c.m.}}$.

The differences between the barriers reflect primarily
the difference between the values for $a_{\text{surf}}$ of these 
parametrizations. As has been pointed out earlier in Ref.~\cite{(Ben00f)},
for parameters sets that are adjusted like ours without an explicit 
constraint on deformation properties, $a_{\text{surf}}$ can take 
very different values depending on the scheme for c.m.\ correction chosen 
during the parameter adjustment. In addition, the present study 
indicates that such fit protocols also produce a weak dependence of $a_{\text{surf}}$ on the value chosen for the isoscalar effective mass, 
cf.\ Table~\ref{tab:INM:ANM}.

The barriers obtained with SLy5s1 and SLy7 are very similar to those 
of the new 1T2T(X) parameter sets, as expected from their similar values
for $a_{\text{surf}}$. There are small differences in detail: SLy7 yields
a smaller excitation energy of the fission isomer, whereas SLy5s1 predicts
it at slightly larger deformation, and both SLy7 and SLy5s1 produce 
a slightly wider outer barrier.

The ground-state deformation takes practically the same value of 
$\beta_{20} \simeq 0.3$ for all parameter sets and agrees well with the 
available experimental data~\cite{(Bem73),Zumbro86}, see the more detailed 
comparison in Sec.~\ref{sec:def:actinides}. 
The deformation of the isomer, however, is slightly
different for each parameter set, mainly in dependence of the effective mass, 
but always remains close to $\beta_{20} \simeq 0.85$. 
This will also be analysed in more detail in Sec.~\ref{sec:def:actinides}. 
Depending on the height of the fission barrier, the positions of the inner 
and outer saddle points also move to slightly larger deformations with 
increasing barrier height, as observed before for the SLy5sX 
series~\cite{Ryssens19a}.

Before entering the comparison with data, we recall that the main 
purpose of our new fits discussed here is not the ``best reproduction''
of barriers by itself, which in one way or another should include 
actual information about deformation in the adjustment protocol, 
but the question of how well barriers are reproduced \textit{without}
considering them in the adjustment protocol depending on the 
choices made for the c.m.\ correction and the isoscalar effective mass.
Phrased differently, we want to analyze which global choices make the 
reproduction of fission barriers a fine-tuning problem within an
existing adjustment protocol that will not be in disproportionate
conflict with other constraints.

Concerning the available experimental data for the barrier, we 
recall that some experiments  for double-humped fission barriers 
provide information about the inner and outer barrier heights, 
while others issue information about the higher (``primary'') and 
lower (``secondary'') of the two barriers.

An example for the analysis of fission of \nuc{240}{Pu} induced 
by direct reactions is Ref.~\cite{Back74a}, which yields $5.80 \pm 0.20$~MeV
and $5.45 \pm 0.20$~MeV for the heights of the inner and outer barrier,
respectively. The data evaluation from multiple experiments provided
by the RIPL-3 database~\cite{RIPL3}, however, lists 6.05 and 5.1~MeV for 
the heights of these barriers. A recent multi-nucleon transfer 
experiment finds $6.25 \pm 0.32 \, \text{MeV}$ for the primary 
fission barrier~\cite{Kean19a}. Values for the excitation energy of the $0^+$ 
superdeformed fission isomer also differ; the authors of Ref.~\cite{Hunyadi01a} 
give $(2.25 \pm 0.20) \, \text{MeV}$, while the data evaluation of 
Ref.~\cite{Singh02a} lists a value of 2.8~MeV. The error bars of the 
experimental values displayed on Fig.~\ref{Pu240} cover the range of these
values.

\begin{figure}[t!]
\includegraphics{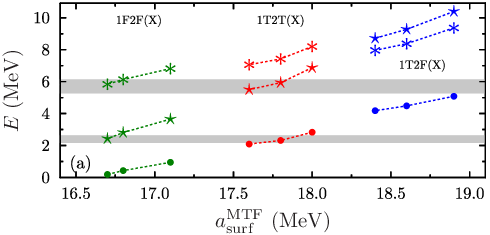}
\includegraphics{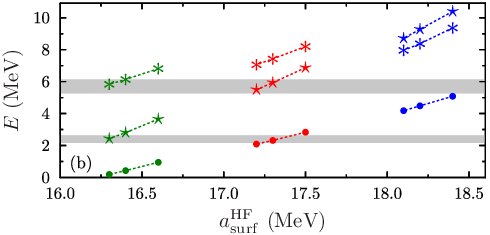}
\includegraphics{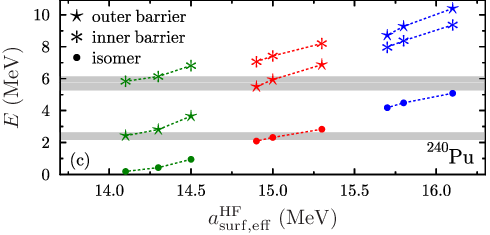}
\caption{\label{pu240:asurf:corr}
Characteristic energies of the fission barrier of \nuc{240}{Pu} as a function
of $a_{\text{surf}}^{\text{MTF}}$ (panel (a)), $a_{\text{surf}}^{\text{HF}}$ (panel (b)),
and $a_{\text{surf,eff}}^{\text{HF}}$(\nuc{240}{Pu}) (panel (c)). Colors indicate
families of parameter sets with same scheme for c.m.\ correction as in 
Fig.~\ref{Pu240}. Markers, however, indicate here the excitation energy
of the fission isomer and the heights of the inner and outer barrier, respectively. 
To guide the eye, lines connect results obtained with the three 
parameter sets with same scheme for c.m.\ correction, but different 
effective mass. Within each
family of parameter sets, $a_{\text{surf}}$ decreases with increasing 
effective mass, see Tab.~\ref{tab:SINM}. As on Fig.~\ref{Pu240},
the experimental data are indicated by horizontal grey
bars, where those for the inner and outer barrier overlap.
}
\end{figure}

From Fig.~\ref{Pu240} it is clear that the 1T2T(X) fits that consider the
full c.m.\ correction give a height of the outer barrier and 
an excitation energy of the isomer that are closest to experiment, 
although neither describes the data perfectly. The parameter sets 
with an elevated effective mass of 0.8 and 0.85 perform slightly
better than the one with $m^*_0/m = 0.7$, but that seems to be 
a particularity  of the fit protocol used for the new parameter 
sets as the calculated barrier obtained with the SLy7 parametrization 
that has an effective mass of $m^*_0/m = 0.69$ is about as close 
to the data.

By contrast, the inner barrier is systematically overestimated by 
all of the 1T2T(X) fits. Its height is only reasonably well described 
by the three 1F2F(X) fits that in turn grossly underestimate the 
excitation energy of the isomer and the height of the outer barrier.

Still, Fig.~\ref{Pu240} confirms the earlier finding that when 
adjusting the parameters of EDFs solely to data on spherical 
ground nuclei and infinite matter, choosing the full c.m.\ correction 
yields more realistic surface properties than choosing the 1F2F 
or 1T2F recipes instead.

Obtaining realistic surface properties for parameter sets of 
1F2F and 1T2F type requires adding information on the surface
energy to the fit protocol. This is exemplified on Fig.~\ref{Pu240} 
by SLy5s1 that produces a barrier of similar quality as the one 
from the 1T2T(0.80) and 1T2T(0.85) and SLy7 parametrizations. Unlike 
these, SLy5s1 is of 1T2F type and had to be constrained during the 
fit to have a realistic value of the surface energy by shifting its
$a_{\text{surf}}^{\text{MTF}}$ value from about 19.0~MeV that it would
naturally acquire to 18~MeV.
Similarly, the parametrizations SkM*~\cite{Bartel82a} (that is of 1T2F type), UNEDF1~\cite{Kortelainen12}, 
and UNEDF2~\cite{Kortelainen14} (both of 1F2F type) that perform similarly well 
for this barrier were also constrained in one way or the 
other to do so.\footnote{Although SkM*~\cite{Bartel82a} is also of 1T2F type, its 
surface energy actually had to be \textit{increased} compared
with the original SkM~\cite{Krivine80} parametrization, see also Ref.~\cite{Jodon16a} 
for a detailed comparison of their $a_{\text{surf}}$ values and the
corresponding fission barrier of \nuc{240}{Pu} calculated in a 
similar manner as done here. The reason is that SkM was adjusted within 
an unusual protocol that focused on nuclear matter properties relevant 
for the description of giant resonances.
} 

The failure of the three 1T2T fits, and also SLy5s1 and SLy7,
to describe simultaneously all of the three characteristic energies
of the barrier of \nuc{240}{Pu} is consistent with the earlier 
findings for nuclei in this mass region~\cite{Bender03r,Jodon16a}. 
Two recent exceptions are BSkG1 and BSkG2~\cite{Ryssens23iib}, which 
describe the inner and outer barriers of \nuc{240}{Pu} similarly well. 

As recalled in Sec.~\ref{sect:asurf}, the surface energy of an EDF
cannot be represented by a unique number, as it has an isospin dependence
and can be determined within different schemes. This poses the question
to which of the various possibilities to characterize surface energy 
the barriers are actually most correlated to. To answer this question, Fig.~\ref{pu240:asurf:corr} displays the excitation 
energy of the fission isomer and the heights of the inner and outer 
barrier of \nuc{240}{Pu} as a function of the isoscalar surface
energy coefficients calculated in MTF ($a_{\text{surf}}^{\text{MTF}}$)
and HF ($a_{\text{surf}}^{\text{HF}}$) approximation, as well as
the effective isospin-dependent surface energy coefficient
$a_{\text{surf,eff}}^{\text{HF}}$ calculated in HF approximation.

The different range of $a_{\text{surf}}$ values over which the 
three parameter sets with same c.m.\ correction scheme are spread
in each of the panels of Fig.~\ref{pu240:asurf:corr} illustrates 
again that the difference between $a_{\text{surf}}^{\text{MTF}}$ 
and $a_{\text{surf}}^{\text{HF}}$ slightly depends on effective 
mass as a consequence of the approximations made in the MTF scheme, 
and that also the surface symmetry energy coefficient that 
enters $a_{\text{surf,eff}}^{\text{HF}}$ takes a slightly different 
value at each effective mass. This change in spread has the 
consequence that the slope of the line connecting results obtained
with the three parameter sets with same c.m.\ correction is different 
in each of the three panels.

When comparing results obtained with parameter sets with 
different effective mass for a given choice of c.m.\ correction, i.e.\ the
data that are plotted in same color on Fig.~\ref{pu240:asurf:corr}, one finds 
in most cases a nearly linear correlation between the calculated 
characteristic energies of the barrier and the respective 
surface energy coefficient. For none of the three choices of surface 
energy coefficient, however, the calculated characteristic energies 
on Fig.~\ref{pu240:asurf:corr} fall near a unique straight line when 
comparing all nine parametrizations from the different families of fits,
i.e.\ the data plotted in different color. Instead, 
there always is an offset when going from one family of parametrizations
to the next. The sign of this offset is also not universal. For the 
excitation energy of the isomer, extrapolating the results from
a family with overall low $a_{\text{surf}}$ to higher $a_{\text{surf}}$ 
will underestimate the results obtained from parameter sets that 
actually have larger $a_{\text{surf}}$. For the height of the 
inner barrier the opposite happens: extrapolating values obtained
with parameter sets that use the same c.m.\ correction to higher 
$a_{\text{surf}}$ will overestimate the barrier height actually 
found for the other families of fits. For the outer barrier 
height, the sign of the offset is even different when plotting 
the values as a function of $a_{\text{surf}}^{\text{HF}}$ or
as a function of either $a_{\text{surf}}^{\text{MTF}}$ or 
$a_{\text{surf,eff}}^{\text{HF}}$. The offsets remain 
comparatively small and do not prevent using any of these 
correlations to adjust a suitable value of $a_{\text{surf}}$ 
as an alternative to the adjustment of actual fission barriers.
However, as already suspected in Ref.~\cite{Jodon16a} based on a
set of parametrizations that was much more limited with respect 
to the choices for c.m.\ correction and effective mass, one can expect
a nearly linear correlation between $a_{\text{surf}}$ and 
deformation energies only when the fundamental choices made 
for the form of the EDF and the fit protocol are the same.
This is of course not surprising as the very origin of the
complicated topography of a fission barrier like the one 
of Fig.~\ref{Pu240} is generated by the variation of shell 
effects that are not directly influenced by $a_{\text{surf}}$, 
but depend sensitively on many of the other choices made
when parametrizing an EDF. 

Shell effects are not the only possible source of such differences.
There are also other contributions to the deformation energy that are not
represented by the surface energy coefficient and therefore can spoil the
correlation between these quantities. One of these is the 
deformation dependence of the pairing correlation energy, i.e.\ the 
energy difference between a HF and a HFB calculation of a nucleus
at given deformation. This energy changes along the fission path as
a nucleus' ground state and fission isomer correspond to deformations 
where pairing correlations are weak because of the low level density 
around the Fermi surface, whereas the saddle points correspond to regions
where pairing correlations are strong because of a large level density 
around the Fermi surface. Assuming that for a given nucleus the size 
of the pairing correlation energy scales with pairing strength, the need 
to adjust the pairing strength separately for parametrizations with 
different effective mass can generate a systematic difference between 
parameter sets with different $m^*_0/m$. This would introduce an effective
mass-dependence of the characteristic energies from parametrizations 
with different effective mass within a series with given scheme for 
c.m.\ correction, and thereby misalign the trends when comparing series 
with different scheme for c.m.\ correction. While such misalignments are seen on
Fig.~\ref{pu240:asurf:corr}, it is unlikely that the pairing correlation 
energy is their main source: for the reasons already mentioned, its effect 
on the excitation energy of the fission isomer should be smaller than 
its effect on the barrier heights, which is not the case for the differences
seen on the figure.

Another contribution for the offsets visible on Fig.~\ref{pu240:asurf:corr}
is the deformation dependence of the c.m.\ correction energy $E_{\text{c.m.}}$ 
displayed in the lower panel of Fig.~\ref{Pu240}. Indeed, $E_{\text{c.m.}}$
does not contribute to the calculation of the surface energy 
of infinite matter, such that its mass- and deformation 
dependence is not represented by $a_{\text{surf}}$ and 
$a_{\text{ssym}}$. Comparing with the 1F2F case where the c.m.\ correction
energy is zero by construction, the values for the excitation energy 
of the fission isomer are pushed up by about 200 keV by this effect 
for parameter sets of 1T2F type, whereas they are pulled down a a few 
tens of keV for parameter sets of 1T2T type. Again, this effect cannot
be the major source for the observed offsets in 
Fig.~\ref{pu240:asurf:corr}, as it is too small in 
absolute size and also cannot explain the relative sign in all cases.

%
%
\subsection{Energy landscape of $^{180}$Hg}
\label{sect:Hg180:barrier}

\begin{figure}[t!]
\includegraphics{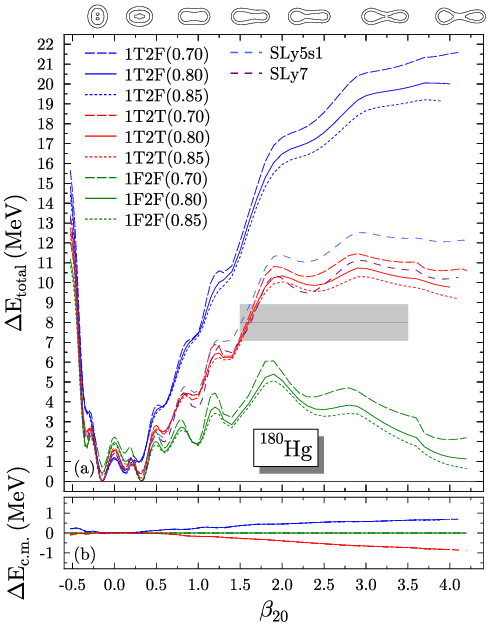}
\caption{\label{Hg180}
Deformation energy curve of $^{180}$Hg  (panel (a)) and change of the
center-of-mass correction (panel (b)) as a function of the quadrupole 
deformation $\beta_{20}$ drawn in the same way as on Fig.~\ref{Pu240}. 
The energy curves end at the deformation at which the calculation jumps 
to a  solution with two separate fragments. 
}
\end{figure}

As a second example we discuss \nuc{180}{Hg}, which is among the most 
neutron-deficient nuclei for which information about the fission barrier 
is available. Because of its much smaller asymmetry $I$, the surface 
symmetry energy is much less important for the barrier of \nuc{180}{Hg} 
than for the one of \nuc{240}{Pu}. 

In addition, this nucleus is situated in a different region of the 
chart of nuclei where shell effects along the fission path are 
very different from those determining the fission path of \nuc{240}{Pu}.
This has several consequences for the energy curves displayed on 
Fig.~\ref{Hg180}. First, \nuc{180}{Hg} exhibits shape 
coexistence of near-degenerate normal-deformed states at low excitation 
energy, one at an oblate deformation of $\beta_{20} \simeq -0.15$,
the other at a prolate deformation of $\beta_{20} \simeq 0.32$. 
Second, model calculations~\cite{Mol2012,Veselsky12a,Warda12a,Li22a} suggest that there is
only one broad barrier, whose saddle point is at very larger deformation, possibly very close 
to the scission point. In fact, the curves on Fig.~\ref{Hg180} end 
where the calculations jump to a solution with two non-identical
fragments. The broad outer barrier follows a reflection-asymmetric 
path beginning at around $\beta_{20} \simeq 1.1$. Like in our earlier 
study of this nucleus with the SLy5sX parametrizations reported in 
Ref.~\cite{Ryssens19a}, we have not found non-axial solutions that 
lower the barrier around the saddle point.

As we are mainly interested in the primary fission barrier of this nucleus,
we have not checked if the various super- and hyperdeformed local 
minima that can be found at intermediate deformations might be 
connected through triaxial shapes that bypass the small barriers 
between them that get particularly pronounced for the parameter 
sets with small surface energy coefficient. For this reason, 
the energy curves shown on Fig.~\ref{Hg180} are for an entirely 
axial fission path.

For the barrier height, comparison with experiment is not entirely 
straightforward as all available data were obtained 
from the observation of $\beta$-delayed fission 
of \nuc{180}{Tl}~\cite{Andreyev10a}, which passes through excited 
states of \nuc{180}{Hg} with negative parity and finite angular 
momentum. The excitation energy of these states is necessarily 
smaller than the $Q$ value for electron capture of \nuc{180}{Tl}, 
$Q_{\text{EC}} (^{180}\text{Tl}) = 10.44~$MeV, which sets an 
upper bound for the fission barrier. The model-dependent analysis 
of the measured probability of $\beta$-delayed fission in that 
nucleus~\cite{Veselsky12a} suggests that the fission barrier has 
a height of about 8.0(9)~MeV, which is the value used in 
Fig.~\ref{Hg180}.

The configuration of \nuc{180}{Hg} for which fission has been observed 
can therefore be expected to have a different structure than the 
ground state for which the fission barrier is calculated. When comparing
theory and experiment, however, we assume that these two barriers are 
the same, as done in the earlier literature on the subject.

The energy curves calculated with the new parametrizations shown 
on Fig.~\ref{Hg180} fall again into three groups according to their 
scheme for c.m.\ correction. Compared to \nuc{240}{Pu}, the differences
are even more dramatic because of the larger range of deformations 
that are probed. This makes it even clearer that the differences in
barrier height cannot be caused by the variation of the c.m.\ correction
itself with deformation. Within each group of parametrizations with
same c.m.\ scheme, the barrier height decreases again with increasing
effective mass, such that the pattern of the energy curves clearly 
follows the sequence of the parametrizations' $a_\text{surf}$ values. 

Comparing the calculated energy curves with data, the 1T2F(X) 
parametrizations again overestimate the barrier height, whereas
the 1F2F(X) underestimate it. While the 1T2T(X) are again closest
to experiment, the calculated barriers are of the same size as 
the upper limit for the barrier from the $Q_{\text{EC}}$ value, 
but overestimate the barrier height as deduced in 
Ref.~\cite{Veselsky12a}.

Like in the case of \nuc{240}{Pu}, SLy7 gives a barrier height 
that falls in between those predicted by the 1T2T(X) parametrizations. By
contrast, SLy5s1 gives a visibly higher barrier than 
the 1T2T(X) although it has a very similar $a_\text{surf,eff}$ value.
This different behavior of SLy5s1 can be explained by 
its different scheme for c.m.\ correction, which for SLy5s1 is of 1T2F type.
As can be seen from panel~(b) of Fig.~\ref{Hg180}, at the 
respective saddle point at $\beta_{20} \simeq 2.8$, the
c.m.\ correction energy of the 1T2F-type parametrizations is about 
500~keV larger than for the ground state, whereas the full 
c.m.\ correction energy of the 1T2T-type  parametrizations becomes 
500~keV smaller. Consequently, the difference in c.m.\ correction 
increases the barrier height of SLy5s1 by about 1~MeV compared 
to SLy7 and the 1T2F(X). 
As the difference in c.m.\ correction grows further beyond the saddle
point, the outmost part of the fission barrier obtained with SLy5s1
is then also somewhat flatter than the one found with any of the 1T2T(X).

As the c.m.\ correction does not enter the calculation 
of the surface energy of semi-infinite matter, the slightly different 
deformation dependence of the c.m.\ correction energy obtained for the 
1F2F(X), 1T2F(X) and 1T2T(X) parametrizations is not accounted for by 
their $a_\text{surf}$ value. For nuclei with a very wide fission 
barrier like \nuc{180}{Hg}, the deformation dependence of the c.m.\ correction 
can therefore make a visible difference for the fission barriers of
parametrizations with same $a_\text{surf}$, but different 
scheme for c.m.\ correction. 
For \nuc{240}{Pu} with its much narrower fission barrier, the 
variation of the c.m.\ correction energy with deformation across the 
barrier is much smaller, such that it does not have a 
visible effect on the fission barrier as seen on Fig.~\ref{Pu240}.

Experimental data consistently points to an oblate shape of the ground
state of this and other \textit{even-even} Hg isotopes in this mass 
region~\cite{Sels19a}, while many EDF parametrizations predict 
a prolate shape for these nuclei instead. Among those that 
do correctly predict an oblate ground state for these nuclei are the fits with low  $a_{\text{surf}}$ 
out of the SLy5sX series such as SLy5s1~\cite{Sels19a}.
This success, however, cannot be attributed to a low $a_{\text{surf}}$ 
value as such. Comparing the new fits, all of the 1T2T(X) parametrizations 
(intermediate $a_{\text{surf}}$) and all of the 1T2F(X) 
(large $a_{\text{surf}}$) predict an oblate ground state,
whereas the 1F2F(X) (low $a_{\text{surf}}$) predict a
prolate ground state. On the other hand, SLy7 predicts a prolate ground state of
\nuc{180}{Hg} although its value for $a_{\text{surf}}$ is
similar to those of the 1T2T(X).

For all of these parameter sets, the energy difference between the 
prolate and oblate minima is at most 1~MeV, and often significantly less. Note that our
calculations also predict a third minimum at small prolate deformation,
that for the 1T2F(X) parametrizations is actually lower in energy than the 
prolate minimum at larger deformation. For the parametrizations with 
large $a_{\text{surf}}$ values out of the SLy5sX series, the weakly
deformed prolate minimum is actually predicted to be the ground state~\cite{Sels19a,Ryssens19a}.

\begin{figure}[t!]
\includegraphics{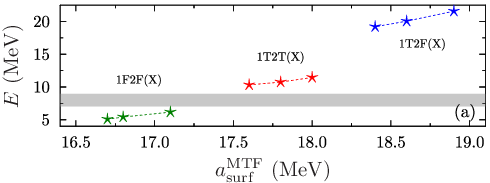}
\includegraphics{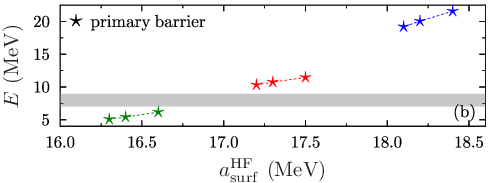}
\includegraphics{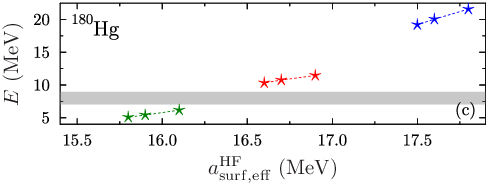}
\caption{\label{hg180:asurf:corr}
Height of the primary fission barrier of \nuc{180}{Hg} as a function
of $a_{\text{surf}}^{\text{MTF}}$ (panel (a)), 
$a_{\text{surf}}^{\text{HF}}$ (panel (b)),
and $a_{\text{surf,eff}}^{\text{HF}}$(\nuc{180}{Hg}) (panel (c)) 
plotted in the same way as on Fig.~\ref{pu240:asurf:corr}.
}
\end{figure}

These minima are generated by shell effects that are related to the 
evolution of the bunching of single-particle levels with deformation
in the Nilsson diagram, and which are more difficult to control 
in a parameter fit than the surface energy. Assuming that these shell 
effects are equal for all of the new fits, then the order of the minima 
obtained with the 1F2F(X), 1T2F(X), and 1T2T(X) parametrizations is 
actually what one would naively expect from the differences between their 
surface  energy coefficients. At small
deformation, the macroscopic deformation energy grows quadratically 
with quadrupole deformation, see Ref.~\cite{Ryssens19a} and references 
therein, such that a state with larger absolute value of $\beta_{20}$ 
looses more macroscopic energy when increasing $a_{\text{surf}}$ than
a state with smaller $\beta_{20}$. That the SLy5sX parametrizations
discussed in Ref.~\cite{Sels19a} do not follow this trend 
indicates that also the ground-state shell effects change significantly 
within this series, which has been illustrated for \nuc{180}{Hg} in 
Ref.~\cite{Ryssens19a}. 
Many traditional Skyrme parametrizations predict a well-deformed 
prolate ground state for even-even Hg isotopes in this mass region, 
and this even in spite of their having large $a_{\text{surf}}$ values that
are comparable to those of the 1T2F(X). This altogether points to
an unresolved fine-tuning problem of shell effects and indicates that
finding the expected $a_{\text{surf}}$-dependence of the energy difference
between the coexisting shapes in \nuc{180}{Hg} for our new fits might
be fortuitous.

We note that Fig.~\ref{Hg180} indicates that the relative energy between the 
various normal-deformed minima of \nuc{180}{Hg} does not show any 
significant dependence on $m^*_0/m$ for the best fits. This is slightly surprising,
as the size and variation of shell effects could have been affected by the effective mass.

Figure~\ref{hg180:asurf:corr} plots the height of the primary fission
barrier of \nuc{180}{Hg} as a function of $a_{\text{surf}}^{\text{MTF}}$, 
$a_{\text{surf}}^{\text{HF}}$, and 
$a_{\text{surf,eff}}^{\text{HF}}$(\nuc{180}{Hg}). Like in the case 
of \nuc{240}{Pu} displayed on Fig.~\ref{pu240:asurf:corr}, there is a
near-linear correlation of the values obtained with the three parameter 
sets with different effective mass but same scheme for c.m.\ correction for 
all of the choices for $a_{\text{surf}}$, but again the barrier heights do 
not perfectly correlate with any of the choices for $a_{\text{surf}}$ 
across families of parameter sets with different c.m.\ correction. While
the deformation dependence of the c.m.\ correction energy mentioned before 
brings an offset of about 1~MeV to the comparison of the results obtained
with the 1T2T(X) and 1T2F(X) sets, there have to be other contributions 
that are even larger.

%
%

\subsection{Deformation}

%
%
\subsubsection{Normal-deformed ground states of actinides}
\label{sec:def:actinides}

\begin{figure}[t!]
    \includegraphics{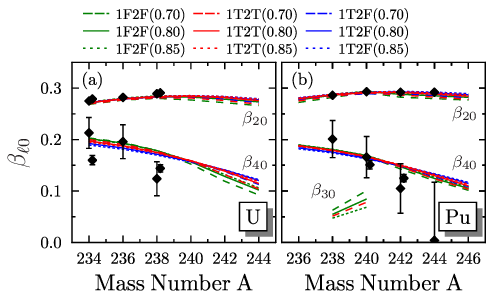}
\caption{\label{betaND} 
Dimensionless quadrupole and hexadecapole deformation of the charge
density distribution of the ground states of even-even U ($Z=92$) and 
Pu ($Z=94$) isotopes compared with experimental data where available, plotted
in the same colors and line styles as on Fig.~\ref{Pu240}.
The error bars of the experimental $\beta_{20}$ values are smaller than the 
markers used to plot them. For the two Pu isotopes for which 
we found reflection-asymmetric ground-state shapes, we also show the 
calculated octupole deformation.
}
\end{figure}

Figure~\ref{betaND} compares the calculated ground-state quadrupole and 
hexadecapole deformations of U ($Z=92$) and Pu ($Z=94$) isotopes with 
the available data\footnote{Note that the $\beta_{2}$ and $\beta_{4}$ 
values given by these references are surface deformations that are 
not equivalent to the volume deformations of Eq.~\eqref{eq:betalm}. 
The experimental $\beta_{\ell 0}$ values used for Fig.~\ref{betaND} 
were obtained from converting the Cartesian quadrupole and hexadecapole
moments given in these references to spherical multipole moments $Q_{\ell m}$ 
and then applying Eq.~\eqref{eq:betalm}.}
for electric transition moments extracted from $B(E2)$
and $B(E4)$ moments determined either from Coulomb excitation~\cite{(Bem73)}
or the analysis of muonic X rays~\cite{(Zum84),Zumbro86}.
We mention that SLy5s1 and SLy7 give results that on the plot are almost 
indistinguishable from those obtained with 1T2T(0.70) 
and therefore have been omitted from the figure.

It is striking to see almost no difference between the calculated values 
obtained from different parametrizations, indicating that for well-deformed 
nuclei with a unique deep normal-deformed minimum in the energy surface 
the ground state deformation is solely determined by the deformation 
dependence of shell effects, but independent on the macroscopic surface
energy. In addition, the experimental $\beta_{20}$ values are almost
perfectly reproduced by all parametrizations. Note that $\beta_{20}$
and $\beta_{40}$ follow a different trend when moving across a major 
shell, with the hexadecapole moment changing sign at about midshell,
which can be understood from the spatial distribution of the single-particle
wave functions that are successively filled, see Refs.~\cite{Bertsch68a,Janecke81a}.
The actinide nuclei for which data are available are located close to 
the region where this happens. Within their large error bars, 
the experimental $\beta_{40}$ values are fairly reproduced, although 
the calculated values tend to decrease too slowly with mass number. 
Although there is a modest spread in the predictions of different self-consistent 
models for hexadecapole deformation in this region, this mismatch between the 
calculated and experimental trend  with mass number seems to be a consistent 
feature of all models that have been used to study this observable, as 
first discussed in  Ref.~\cite{(Lib82)} and shown explicitly in the case of 
$^{238}$U for 21 different parametrizations of Skyrme's EDF in 
Ref.~\cite{ryssens23c}. Given this indication from different models and 
the inherent difficulties of the experimental determination of hexadecapole 
deformation, it seems worthwhile to revisit this region with modern technology. 
For $^{238}$U in particular, such experimental
information would be complementary to information that might be gleaned from 
ultra-relativistic heavy-ion collisions of this 
nucleus~\cite{magdy2023}.

As mentioned when discussing the fission barrier of \nuc{240}{Pu}
in Sec.~\ref{sect:Pu240:barrier}, for a few parametrizations  we find
an octupole-deformed ground state for this nucleus that is accompanied by 
a very small energy gain of at most 200~keV for 1F2F(0.70). 
That the deformation energy surface of \nuc{240}{Pu} is soft against
octupole deformation has been noticed before
\cite{robledo_2012,Agbemava16a}, indicating the
possibility of dynamical octupole correlations that would explain
the experimentally observed low-lying negative-parity band whose
levels decay to states in the ground-state band via strong $E1$ transitions
\cite{Wang09a,Spieker18a}. For the same four parametrizations, 
1F2F(0.70), 1F2F(0.80), 1T2F(0.85), and 1T2T(0.70), we also find a
shallow octupole-deformed minimum for \nuc{238}{Pu}, again with an 
energy gain that does not exceed a few tens of keV. The $\beta_{30}$
deformation of these isotopes are also displayed on Fig.~\ref{betaND}.
Compared with the lowest reflection-symmetric configuration of these isotopes,
the quadrupole and hexadecapole deformations do not change by an 
amount that can be resolved on the figure. No octupole-deformed
minima are found for the heavier plutonium isotopes or any of the 
uranium isotopes displayed on Fig.~\ref{betaND}.

\begin{figure}[t!]
    \includegraphics{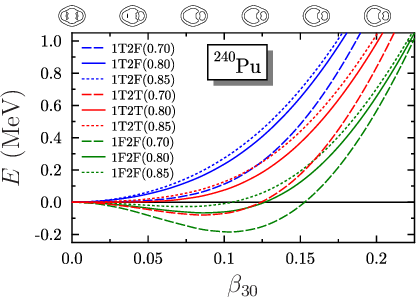}
\caption{\label{pubeta30} 
Energy of the normal-deformed configuration of \nuc{240}{Pu} as a function
of dimensionless octupole deformation $\beta_{30}$, plotted
in the same colors and line styles as on Fig.~\ref{Pu240}.
The inserts on top of the figure indicate the typical evolution of 
shapes along the energy curve.
}
\end{figure}

For the more neutron-deficient nuclei displayed on this figure,
the presence or absence of an octupole-deformed minimum results from 
a small change in the softness of the deformation energy surface with
respect to $\beta_{30}$, as  is illustrated by Fig.~\ref{pubeta30} 
for \nuc{240}{Pu}. The pattern of 
differences between the deformation energy curves is clearly correlated 
with the effective mass and the scheme for c.m.\ correction employed: the energy curve 
becomes stiffer when going from the 1F2F(X) to the 1T2T(X) and then to the 1T2F(X), 
reflecting the global dependence of $a_{\text{surf}}$ on the scheme of
c.m.\ correction. For parametrizations with the same scheme for c.m.\ correction, 
it is, however, the one with the \textit{smallest} effective mass that is the softest 
against octupole deformation. Therefore, the sequence of energy curves
is not directly determined by $a_{\text{surf}}$, as for a given c.m.\ scheme
it is the parametrization with the \textit{largest} $m^*_0/m$ that has the 
smallest $a_{\text{surf}}$ value (see Table~\ref{tab:SINM}). This points to an 
important role of the effective mass for the variation of shell effects
with deformation that generate the octupole-deformed minima in this mass
region. Comparing the parametrizations that generate an octupole-deformed 
minimum for this nucleus, the size of octupole deformation at the minimum 
and the energy gain are clearly correlated.

We mention that a similar pattern, but with much larger gain in deformation 
energy from octupole deformation, is also found for \nuc{222}{Ra}~\cite{dacosta_22},
a nucleus for which empirical data point to static octupole deformation.
Altogether, Fig.~\ref{pubeta30} confirms the finding of Ref.~\cite{Ryssens19a}
that it is more likely to find static octupole deformation for parametrizations
of Skyrme's EDF with low surface energy coefficient. The present study points
to a second nuclear property of EDFs that amplifies such exotic deformation 
modes, which is a small effective mass as suspected in 
Ref.~\cite{PhysRevC.76.034317}.

While our finding of octupole-deformed minima for a few plutonium 
isotopes indicates that lowering $a_{\text{surf}}$ of a parametrization 
significantly increases the likelihood of shape transitions that 
involve exotic shape degrees of freedom in self-consistent mean-field
calculations, the significance of the actual octupole-deformed 
minima for the interpretation of experimental data is less clear.
The minima are too shallow to interpret the nuclei for which they
are found as rigid octupole-deformed rotors, which also would be 
incompatible with experimental data for the observed states at low spin. 
Still, that fluctuations on octupole degrees of freedom 
might play a significant role for \nuc{240}{Pu} is evident when comparing 
Fig.~\ref{pubeta30} with Fig.~\ref{Pu240}: around the ground state, 
the energy surface of \nuc{240}{Pu} is much softer with respect to 
octupole deformation than with respect to quadrupole deformation for 
all nine of the ``best fit'' parametrizations, irrespective of their 
predicting an octupole-deformed minimum or not.

%
%
\subsubsection{Superdeformed fission isomers of actinides}

\begin{figure}[t!]
    \includegraphics{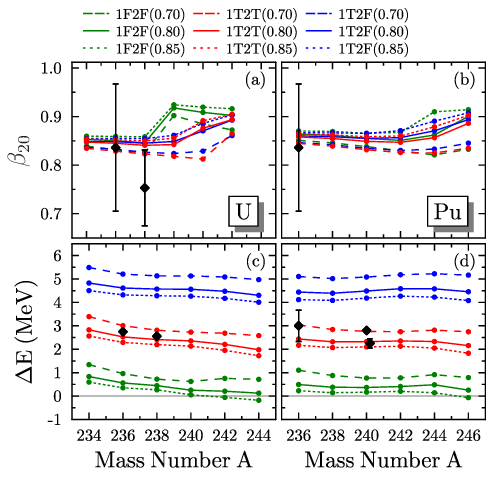}
\caption{\label{betaSDAct}
Charge quadrupole deformation (panels (a) and (b)) and excitation energy 
(panels (c) and (d)) of the $0^+$ fission isomers of even-even U 
(panels (a) and (c)) and Pu (panels (b) and (d)) isotopes.
Colors and line styles are the same as in the previous figures.
}
\end{figure}

Figure~\ref{betaSDAct} compares calculated values for the excitation energy
and quadrupole deformation of superdeformed (SD) fission isomers of U 
($Z=92$) and Pu ($Z=94$) isotopes with the available data. As done earlier 
in Ref.~\cite{Nikolov11a}, we limit the comparison to data for isomers 
that could be identified as $0^+$ bandheads.

The excitation energies of the fission isomers of the uranium isotopes 
are taken from Ref.~\cite{Garg23a}, the energy of the state with 37.4~ps 
lifetime of \nuc{236}{Pu} from~\cite{Singh02a}, and the energy of the 
isomer of \nuc{240}{Pu} from Refs.~\cite{Hunyadi01a,Singh02a}, see 
also Sec.~\ref{sect:Pu240:barrier}.

The experimental $\beta_2$ values were obtained converting the Cartesian 
charge quadrupole moments $Q_0$ listed in Ref.~\cite{Metag80a} to 
spherical quadrupole moments $Q_{20} = \sqrt{5/(16 \pi)} \, Q_0$ first 
and then applying Eq.~\eqref{eq:betalm}.

As could be expected from the discussion of the fission barrier of
\nuc{240}{Pu}, the 1T2F(X) parametrizations grossly overestimate the known
excitation energies of fission isomers. This performance is similar to 
almost all other Skyrme parametrizations that use the 1T2F recipe and that are not fine-tuned to 
describe highly-deformed states. The 1F2F(X) on the other hand 
grossly underestimate this energy, so much so that SD minima become the 
global minima for some heavy actinide nuclei. The isomer excitation energies
predicted by the 1T2T(X) are compatible with available data, further confirming that choosing the 1T2T(X) recipe 
for c.m.\ correction automatically leads to quite realistic, although not 
completely perfect, surface properties. 

Taking into account the huge error bars on $\beta_{20}$, one can consider that all new parameter sets agree with data for 
the quadrupole deformation. Unlike the case of the normal-deformed minima,
the calculated values do not fall on top of each other which indicates 
that there is some variation in the shell structure predicted in the second well.

We have checked that for all parametrizations the SD minimum is stable 
with respect to non-axial and reflection-asymmetric deformations.

%
%
\subsubsection{Superdeformed states of Hg and Pb isotopes}

\begin{figure}[t!]
\includegraphics{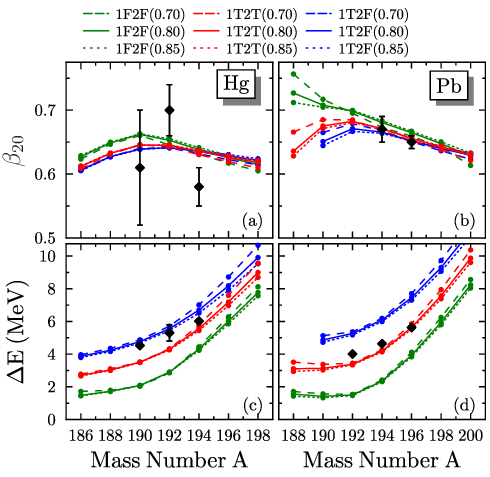}
    \caption{\label{betaSD}
     Charge quadrupole deformation (panels (a) and (b)) and excitation energy 
     (panels (c) and (d)) of the hypothetical (see text) $0^+$ bandheads 
     of the superdeformed rotational bands of even-even Hg 
     (panels (a) and (c)) and Pb (panels (b) and (d)) isotopes. 
     Colors and line styles are the same as in the previous figures.
            }
\end{figure}

Figure~\ref{betaSD} compares predictions for the excitation energy and
charge quadrupole moment of the $0^+$ bandheads of SD rotational of 
even-even neutron-deficient Hg ($Z=80$) and Pb ($Z=82$) isotopes with 
available data.

Again, the experimental quadrupole deformations $\beta_2$ have been deduced 
from Cartesian transition quadrupole moments\footnote{Note that the 
$\beta_2$ values given in the same Table of Ref.~\cite{Singh02a} are 
surface deformations that are not equivalent to the volume 
deformations defined through Eq.~\eqref{eq:betalm}, see 
Ref.~\cite{Ryssens19a} and references therein.}
$Q_t = \sqrt{16 \pi/5} \, Q_{20}$ listed in Ref.~\cite{Singh02a}. 
These quadrupole moments are obtained from averaging transition
moments between high-spin states built on top of the respective 
band head. 

In general, the calculated $\beta_2$ values of Pb isotopes 
are slightly larger than those of Hg isotopes with same 
neutron number, pointing to a significant role of proton shell 
effects for the SD minimum. For both the Hg and Pb chains, the 
calculated $\beta_2$ take their maximum value at about $N \simeq 110$. 
For Hg isotopes, the calculated $\beta_2$ slightly fall off on both sides,
whereas for Pb isotopes, only the values calculated with the 
1T2F(X) and 1T2T(X) parametrizations follow this trend. 
For the heavier isotopes of both elements with $N \gtrsim 110$, all 
new fits predict similar $\beta_2$ values that fairly reproduce the 
available data that have very large error bars. For the most 
neutron-deficient isotopes, however, the 1F2F(X) systematically 
yield slightly larger values than the fits from the two other 
series. This can possibly be attributed to 1F2F(X)'s $a_{\text{surf}}$ 
values being smallest among all new fits and therefore yielding 
the softest deformation energy surfaces.

Not all Hg and Pb isotopes exhibit a SD minimum, and Fig.~\ref{betaSD}
is limited to the range of neutron numbers for which it is most likely 
to find one. Not all parametrizations predict an SD minimum for the 
same range of neutron numbers, as indicated by the 1T2F(X) for which 
none is found for \nuc{188}{Pb}. That the likelihood of finding a SD
minimum increases with decreasing $a_{\text{surf}}$ has already been 
illustrated by Fig.~\ref{Hg180} for \nuc{180}{Hg}: the flatter the 
deformation energy surface, the more likely it is that local variations 
of shell effects generate local minima.

The experimental data for the excitation energies of the bandheads
of the SD rotational bands in these nuclei are taken 
from Refs.~\cite{(Wil10),(Koo96),(Lau00)}. The bandheads themselves 
have not been identified in experiment so far; instead, their energy is
estimated from the extrapolation of the excitation energies of high-spin 
levels in the rotational band built on top of them.

Going towards more neutron-deficient isotopes, the excitation 
energy $\Delta E$ of the calculated SD bandheads first decreases rapidly 
and then levels out. As can be expected from their $a_{\text{surf}}$
values, the curves obtained from the 1F2F(X), 1T2T(X) and 1T2F(X)
are almost parallel, with an offset of about 1.5~MeV when going from
one series to the next. The available experimental data, which are all in 
the region where the slope of the calculated $\Delta E$ starts to level out,
decrease slightly less quickly than the calculated ones. None of the new 
fits describes simultaneously the data for Hg and Pb isotopes: While the 
$\Delta E$ of Pb isotopes are reasonably well described by the 1T2T(X) 
-- which are also those that performed best for all other deformation 
energies discussed so far -- the same parametrizations visibly underestimate
the $\Delta E$ of the Hg isotopes. By contrast, the 1T2F(X) with their 
larger $a_{\text{surf}}$ fairly describe these data.

This discrepancy in performance for the $\Delta E$ of adjacent Hg and Pb 
isotopes is likely to be a deficiency in the description of the relative 
size of shell effects in the various minima of Hg isotopes.
The same flaw has been found for the SLy5sX series in 
Ref.~\cite{Ryssens19a}: SLy5s1 fairly describes the ground state and 
fission barrier of \nuc{180}{Hg}, the fission barrier of \nuc{240}{Pu}
and the SD bandheads of Pb isotopes, but underestimates the SD band heads
of Hg isotopes by a similar amount to what is found here for the 1T2T(X). 
That the $\Delta E$ of Hg and Pb isotopes is not simultaenously described
by widely-used parametrizations of the Skyrme EDF had already been pointed
out earlier in Ref.~\cite{Heenen98a}.


\subsubsection{Shape coexistence of even-even Hg isotopes at normal deformation}

As we mentioned already when discussing the fission barrier of \nuc{180}{Hg} on Fig.~\ref{Hg180}, 
there is experimental evidence that the ground states of \textit{even-even} Hg 
isotopes below $N \simeq 120$ are the weakly oblate-deformed band heads 
of a collective rotational band, which at least for isotopes
between $100 \leq N \leq 110$ coexists with an excited prolate rotational 
band that has much larger a moment of inertia~\cite{Heyde11a}. The excitation 
energy of the $0^+$ state interpreted as the prolate band head roughly 
follows a parabolic trend with $A$~\cite{Heyde11a}, taking its minimal 
value of 328~keV for $A=182$.
For three of the intermediate \textit{odd-mass} Hg isotopes with $101 \leq N \leq 105$
around the minimum of this parabolic trend, however, a prolate state becomes 
the ground state, which leads to an anomalous odd-even staggering of charge radii 
\cite{Heyde11a,Sels19a} in this mass region.

Nuclear EDF methods in general reproduce the coexistence of oblate 
and prolate states in this mass region. What most nuclear EDF
methods fail to reproduce is the relative order and mass dependence 
of the energy difference between the prolate and oblate states~\cite{Sels19a}.

Energy curves from calculations
limited to axial symmetry often exhibit an additional third weakly-deformed 
prolate minimum that, however, might turn out to be a saddle point 
when considering more general non-axial shapes. When allowing for non-axial
shapes, some of the calculated well-deformed prolate states become slightly triaxial 
for each of our best fit parametrizations. In most cases this concerns the two 
heaviest even-even isotopes for which such minimum is found, for the 1F2F(X) even 
the heaviest three, but for 1T2F(0.85) only the heaviest one. In any event, these are not  
the same isotopes for each parameter set. For \nuc{180}{Hg} discussed earlier,
and with energy curves shown on 
Fig.~\ref{Hg180}, only 1T2F(0.80) predicts slight triaxiality of its quite highly excited 
prolate state, which is accompanied by an energy gain of 50~keV. The triaxiality angle 
$\gamma$ typically takes values between 7 and 15 degrees, and in most cases increases 
with $A$. Simultaneously, the energy gain from triaxiality also increases, taking 
values of up to about 350~keV for the last isotope for which such minimum is found. 
Altogether, this situation is quite different from the case of normal-deformed prolate 
states of Pu isotopes (shown on Fig.~\ref{betaND}) for which the occurence of octupole deformation 
is correlated to $a_{\text{surf}}$ and $m^*_0/m$.

\begin{figure}[t!]
\includegraphics{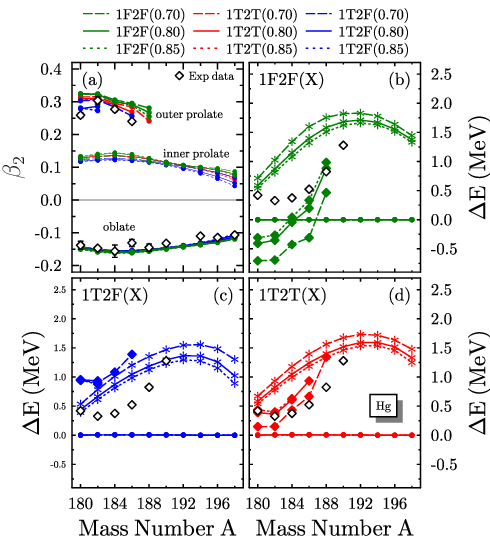}
    \caption{\label{hg:corx}
             Shape coexistence at normal deformation in even-even
             neutron-deficient Hg isotopes. Panel (a): 
             calculated dimensionless 
             quadrupole charge deformation $\beta_{2}$ 
             of the three minima compared with experimental data
             where available (see text). Panels (b), (c), (d): energy of the weakly
             and strongly deformed prolate states relative to the oblate
             state shown separately for the 1F2F(X), 1T2F(X) and 1T2T(X)
             parametrizations.
             Colors and line styles are the same as in the previous figures. 
             }
\end{figure}

Panel~(a) of Fig.~\ref{hg:corx} compares the total quadrupole deformation 
$\beta_{2} = \sqrt{\beta_{20}^2 + 2 \, \beta_{22}^2}$  of the 
calculated minima found with the available data. We multiplied
the $\beta_{2}$ values of oblate states by a minus sign for better separation of the curves.
Experimental data for the absolute $\beta_{2}$ values of the oblate states 
as determined from $B(E2,2^+_1 \to 0^+_1)$ values through the rigid rotor model
are taken from the NUDAT database~\cite{NUDAT}. There are indications that 
the low-lying coexisting states of some of these Hg isotopes are strongly 
mixed~\cite{Wrzosek19a}; 
therefore, the rigid-rotor model cannot be expected to perfectly describe 
these transitions. Still, the calculated $\beta_{20}$ of the oblate states 
agree well with these data. In particular, they reproduce well the slightly parabolic 
trend with $A$.

The experimental data for the absolute $\beta_{2}$ values of the prolate 
states are deduced from the $B(E2,6^+_1 \to 4^+_1)$ values measured in the
experiments reported in Ref.~\cite{Grahn09a,Gaffney14a}, again through 
the rigid rotor model.\footnote{Note that the $\beta_2$ reported in 
Ref.~\cite{Grahn09a} are again surface deformations, not volume deformations
as plotted on Fig.~\ref{hg:corx}.}
Because of their large moment of inertia,
these states can be attributed to the rotational bands built on the prolate 
state of the respective nucleus. They are yrast for all nuclei for which 
there are data and expected to be less mixed with the oblate states than 
the lower-lying ones in this band, which makes the extraction of the 
transition quadrupole moment through the rigid-rotor model more reliable.
The overall size and $A$ dependence of the deformation of the prolate 
states is  also well reproduced, although calculated values fall off 
less quickly with increasing $A$. This may simply point to 
the inadequacy of the mean-field ansatz to model the
complex structure of these states~\cite{Wrzosek19a,Yao13a}.

The other three panels of Fig.~\ref{hg:corx} compare the energy of the 
coexisting normal-deformed minima to the energy of the oblate state with the 
available data. As we saw already in the discussion of the fission barrier 
of \nuc{180}{Hg} represented on Fig.~\ref{Hg180}, depending on the 
choices made for the scheme of c.m.\ correction the new fits make very 
different predictions for shape coexistence at normal deformation
of \nuc{180}{Hg}, which is a direct consequence of the very different
$a_{\text{surf}}$ values of these fits. 

For the 1T2F(X) [Fig.~\ref{hg:corx}(c)] 
that have the largest $a_{\text{surf}}$, the ground state is oblate and
the excitation energy of the well-deformed prolate states is grossly 
overestimated. Interestingly, the calculations do not find such prolate 
minimum for all isotopes for which a prolate rotational band is known.
The second prolate minimum at smaller deformation remains well above the 
oblate state for all mass numbers. 

For the 1T2T(X) [Fig.~\ref{hg:corx}(d)] that have intermediate  $a_{\text{surf}}$ values,
the ground state is also oblate, but now the well-deformed prolate states 
are at about the correct energy. Note that, without the additional energy 
gain from triaxial deformation, the excitation energy of the heavier 
isotopes would be further off the data.
Unlike the case of the 1T2F(X), there is a visible effective-mass dependence 
of the excitation energy of the prolate state: reducing $m^*_0/m$ also 
lowers the excitation energy. 

For the 1F2F(X) that have the lowest $a_{\text{surf}}$ values, the
oblate and well-deformed prolate structures cross in energy, such that the 
lightest of the Hg isotopes shown in  Fig.~\ref{hg:corx} have a prolate ground state.
For the 1F2F(X), the effective-mass dependence of the excitation energy of the 
well-deformed prolate minimum is even more pronounced than for the 1T2T(X).

Comparing the three series, there is also a striking difference that concerns
the isotopes for which a well-deformed prolate minimum is found: the range in $A$ 
is smallest for the 1T2F(X) and largest for the 1F2F(X) parametrizations. More specifically, going from 
1T2F(X) to 1T2T(X) and then to 1F2F(X) at a given effective mass, the heaviest 
isotope for which a well-deformed prolate minimum is found is in most cases 
pushed two mass units further up. There also is an effective-mass dependence: 
for 1T2T(0.70) and 1T2F(0.70), such minima are still found
two mass units further up than for the parameter sets with larger effective
mass from the same series.

In all cases, the excitation energy of the prolate bandhead varies too quickly 
with mass number. Finding a well-deformed prolate minimum also seems  to be
correlated to its excitation energy: there are no such minima found at
more than about 1.2~MeV above the oblate state.

As already noted when discussing Fig.~\ref{Hg180}, the differences in 
relative energy between the weakly-deformed oblate and well-deformed prolate 
minimum when comparing the 1F2F(X), 1T2T(X) and 1T2F(X) directly reflect the 
differences in their $a_{\text{surf}}$ values: increasing $a_{\text{surf}}$ leads
to a larger loss in binding energy for the minimum at larger deformation.

The 1T2T(X) perform best for this phenomenon, confirming again
that adjusting a parametrization of the Skyrme EDF at NLO with the full c.m.\
correction leads to quite realistic deformation properties, even when no 
information on deformed nuclei enters the adjustment protocol.

We mention in passing that for obvious reasons the 1T2T(X) are the only 
parametrizations out of the new fits that produce an anomalous odd-even
staggering of the light Hg isotopes~\cite{dacosta_22}; like in the case of SLy5s1 discussed
in Ref.~\cite{Sels19a}, however, the phenomenon is not predicted for exactly
the same mass range at which it is observed experimentally.

%
%

\section{Summary, Conclusions and Outlook}
\label{sec:summary}

We investigated the impact of choices made for the scheme of center-of-mass correction 
and the isoscalar effective mass $m^*_0/m$ on the resulting surface properties of nuclear 
EDF through a series of dedicated fits of parameter sets of the widely-used standard NLO 
form of the Skyrme EDF.

To this aim we first constructed nine series of parametrizations that differ in
their scheme for c.m.\ correction; i.e.\ none (1F2F), one-body term only (1T2F) and 
full one and two-body correction (1T2T), and in their isoscalar mass, i.e.\ $m^*_0/m = 0.70$,
$0.80$, and $0.85$. Adding a constraint on the surface energy coefficient 
$a_{\text{surf}}^{\text{MTF}}$ calculated using the MTF approximation
to a fit protocol that otherwise only constrains properties of doubly-magic spherical nuclei 
and properties of infinite matter, we constructed a set of parametrizations for each
combination of c.m.\ correction strategy and $m^*_0/m$ that covers the wide range of 
$a_{\text{surf}}^{\text{MTF}}$.
The main observations and conclusions drawn from the analysis of these parametrizations are:
\begin{itemize}
\item 
The value of the penalty function of the adjustment 
protocol of the converged parameter fits varies strongly with $a_{\text{surf}}^{\text{MTF}}$ 
within each of the nine series of fits.

\item 
The optimal value for $a_{\text{surf}}^{\text{MTF}}$ that gives the smallest value for 
the penalty function within a given series of fits depends strongly on the choice made for the 
scheme of c.m.\ correction in the EDF, as has been deduced earlier~\cite{(Ben00f)} in a 
much more limited study.  
In addition, there is a mild dependence of value for $a_{\text{surf}}^{\text{MTF}}$ 
that minimizes the penalty function on the isoscalar effective mass $m^*/m$.

\item 
We find strong correlations between almost all properties of infinite nuclear matter
and the constrained value for $a_{\text{surf}}^{\text{MTF}}$. The origin of these correlations 
is probably threefold. 
First, there is a physics reason that can be qualitatively explained in the 
liquid-drop model: varying $a_{\text{surf}}$ changes the contribution from the 
surface energy to the total binding energy of finite nuclei. To achieve a similar 
description of binding energies with different values of $a_{\text{surf}}$,
other contributions to the LDM energy have to absorb the change in surface 
energy through a change of their coefficients.
Second, there is a limitation of the standard Skyrme EDF: 
the number of its coupling constants is smaller than the number of relevant nuclear
matter properties, which introduces an inevitable correlation between virtually 
all nuclear matter properties and the size of $a_{\text{surf}}$.
As we cannot expect that the standard Skyrme EDF provides perfect modeling of 
nuclear systems and covers all physical degrees of freedom, this limitation
of the Skyrme EDF introduces unphysical interdependencies between nuclear 
matter properties.
Third, there is an accidental interconnection between the scheme chosen for the c.m.\
correction and the properties of nuclear matter properties. Although the c.m.\ correction
itself does not contribute to the properties of infinite and semi-infinite matter,
using different schemes during the parameter adjustment produces parametrizations with
different nuclear matter properties as the other contributions to the total binding energy
have to absorb the differences between the resulting c.m.\ correction energy.

\item 
It is likely that similar correlations between the surface and surface symmetry energy 
will be found when constructing series of parametrizations with 
varied infinite matter properties.

\item
We  confirm earlier studies~\cite{Jodon16a} that, for NLO Skyrme EDFs, the values for 
$a_{\text{surf}}$ obtained with different schemes to calculate semi-infinite matter, systematically differ by an offset. The MTF approach
systematically gives values that are larger than the HF ones by a few hundreds of keV.
The size of this offset depends on effective mass, which can be attributed to the 
\textit{ansatz} for the kinetic density that is made in the MTF scheme. 
By contrast, values for $a_{\text{surf}}$ obtained from the ETF approximation are 
systematically smaller than the HF ones, again by a few hundreds of keV. This time however, 
the difference between the $a_{\text{surf}}$ slowly increases with their absolute size 
with a mild effective mass dependence. This confirms that the MTF value can serve
as an efficient tool to constrain the isoscalar surface energy coefficient $a_{\text{surf}}$ 
in a parameter fit. Unfortunately, extending the MTF scheme to asymmetric matter is not
straightforward~\cite{Krivine83} and requires additional approximations when
$a_{\text{ssym}}$ is also to be constrained. In addition, the MTF \text{ansatz} is 
specifically tailored for the Skyrme NLO EDFs
and cannot be applied to Skyrme EDFs of higher order in gradients
that require the set-up of an alternative scheme~\cite{Proust22a}.

\end{itemize}
In a second step, we constructed nine fits without constraint on $a_\text{surf}^\text{MTF}$
that each represent the ``best fit'' for a given combination of choices for c.m.\ correction and 
isoscalar effective mass in the sense that they correspond to the minima of the penalty 
function of our adjustment protocol for each of the nine series constructed with a 
constraint on $a_\text{surf}^\text{MTF}$. Like the majority of parametrizations of the
Skyrme EDF, the adjustment protocol of these nine ``best fits'' only considers properties
of spherical nuclei and infinite nuclear matter, but no information on deformation properties
of finite nuclei. With this, these parametrizations are representative of the consequences
of the choices made for the scheme of c.m.\ correction and the effective mass on the surface
energy of Skyrme EDFs at NLO. The main observations and conclusions from our analysis of
their nuclear matter properties are:
\begin{itemize}
\item 
As a consequence of the correlations between $a_\text{surf}$ and properties
of infinite matter, the INM properties of the nine best fits systematically differ
and this even in spite of some of them being constrained by the adjustment protocol.

\item 
Most importantly, the nine ``best fits'' have systematically different values for the surface 
energy coefficient.
First, there is a clear dependence on the scheme for c.m.\ correction: for the parameter sets 
using the 1T2F scheme, $a_{\text{surf}}$ is almost 1~MeV larger than for parameter sets employing the 
full 1T2T scheme, whereas for parametrizations using the 1F2F scheme it is about 1~MeV
smaller. This effect has already been identified for the difference between fits of
1T2F and 1T2T type before~\cite{(Ben00f)}. Our results demonstrate that something similar, but in the 
opposite direction, happens for fits of 1F2F type. 
On top of that, we also observe a mild dependence of
$a_\text{surf}$ on effective mass, at least in our fit protocol.

\end{itemize}
The main observations and conclusions from our analysis 
of the deformation energies of finite nuclei obtained with these nine parametrizations are:
\begin{itemize}
\item 
For all examples we studied, the calculated energy differences between two configurations 
in the same given nucleus scale roughly with the surface energy 
coefficient $a_\text{surf}$ of the parametrization used. For some observables, but not all, there is an additional
dependence on effective mass.
The former of these two correlations can be expected from the deformation
dependence of surface energy in the liquid-drop model, whereas
the latter results from a deformation-dependence of shell effects.

\item At small deformation, which means regions where the macroscopic liquid-drop energy
is slowly varying with deformation, the actual deformation at which mean-field minima
are found is rather insensitive to the value of $a_{\text{surf}}$. 
By contrast, the deformation of highly deformed excited states situated on the 
flank of a high fission barrier where the macroscopic energy varies quickly
shows some dependence on $a_{\text{surf}}$. In addition, highly-deformed
minima for some nuclei, are only found for parametrizations with low $a_{\text{surf}}$.
From the point of view of Strutinski's theorem, we attribute this behavior to 
the relative rate at which microscopic and macroscopic contributions contained 
in the EDF change with deformation. The former are determined by variations of
the average density of single-particle levels around the Fermi energy, whereas the
latter roughly increase quadratically with deformation, at least up to the point 
where the nucleus forms a neck. When the shell effects vary quicker than the
macroscopic background, they determine the position of minima in the energy surface.
By contrast, when the macroscopic background varies quicker than shell effects,
then the barriers and minima obtained from the combined contributions move in
deformation or might disappear completely.

\item The 1T2T(X) fits provide the 
best overall agreement with experiment, particularly the parametrizations with elevated effective mass.
Within the uncertainties of the experimental data, the 1T2T(X) parametrizations describe fairly well the 
fission barriers of \nuc{240}{Pu} and \nuc{180}{Hg}, ground-state de\-for\-mation of 
actinides, shape coexistence in neu\-tron-deficient Hg isotopes, and the superdeformed 
states of actinides and Pb isotopes. 
The only clear deficiency  of the 1T2T(X) that we found is their underestimation of the 
excitation energy of the superdeformed bandhead of some Hg isotopes. By contrast, 
the 1F2F(X) systematically underestimate all deformation energy differences, whereas 
the 1T2F(X) almost always overestimate them. 

\item By no means, however, do the 1T2T(X) offer the best possible description of deformation 
energies that can be achieved for a Skyrme NLO EDF. This was not our purpose; instead these parametrizations demonstrate that a  reasonable description of deformation
energies can be achieved without explicitly considering information about 
deformation  energies in the adjustment protocol by simply choosing the 1T2T scheme for
c.m.\ correction. For parametrizations using the 1T2T scheme, the accurate
description of deformation energies becomes a fine-tuning problem. For
parametrizations using the 1T2F or 1F2F schemes on the other hand, the adjustment of deformation 
energies will require a major degradation of other properties. This is consistent 
with the recent BSkG1, BSkG2 \cite{Ryssens23iib}, and BSkG3 \cite{grams23a} 
parametrizations, which all use the 1T2T recipe, achieving an excellent simultaneous 
description of masses, charge radii, fission barriers and nuclear matter properties relevant for nuclear astrophysics in case of BSkG3.

\item 
The differences in deformation energy found between the 1F2F(X), 1T2T(X) and 1T2F(X)
fits is almost independent on the contribution of the c.m.\ correction itself to
the total energy. The slow variation of the c.m.\ correction with deformation 
only makes a visible difference for nuclei with a very wide fission barrier such 
as \nuc{180}{Hg}.

\item Our results confirm the finding of Ref.~\cite{Ryssens19a} that the likelihood of 
finding minima in the energy surface for configurations with exotic shapes 
increases with decreasing surface energy coefficient of the employed parametrizations. 
Our results point in addition to a significant role of the effective mass in this 
respect. This point deserves further study in the future.

\item Our findings explain a number of observations made in the literature about the performance
of Skyrme EDFs regarding nuclear deformation properties. Unfortunately the majority of 
Skyrme EDFs for nuclear structure and nuclear matter studies are still adjusted with 
the numerically less costly 1T2F scheme, which tends to make nuclei too rigid
against deformation. Constructing parametrizations for nuclear dynamics with the 1F2F 
scheme can also pose problems since such strategy tends to make nuclei too soft against deformation
unless surface properties are explicitly tuned during the fit.

\end{itemize}

Our study raises the question to which extent not incorporating other 
quantal effects that cannot be easily described by mean-field modeling based on an EDF might also
be spuriously imprinted on the properties of the EDF's parametrizations. The most immediate suspects are
rotational and vibrational corrections for collective motion but the Wigner energy
might be another~\cite{goriely2002}.

There also is a noteworthy difference between the optimal values for 
$a_{\text{surf}}$ when comparing different types of models. For the Skyrme
EDFs used here, the best description of barriers is achieved for 
$a_{\text{surf}}^{\text{HF}} \simeq 16.4 \, \text{MeV}$ 
in combination with $a_{\text{ssym}}^{\text{HF}} \simeq -46 \, \text{MeV}$.
The surface and surface-symmetry energy coefficients of macroscopic-microscopic 
models, for which they usually are adjusted to fission barriers,
are very different from these values.
The FRLDM model of Ref.~\cite{Moller16a} gives
$a_{\text{surf}} = 21.269461~\text{MeV}$ and
$a_{\text{ssym}} = -50.804~\text{MeV}$,
and the three LDM models fitted in Ref.~\cite{Pomorski03a} 
have $a_{\text{surf}}$ and $a_{\text{ssym}}$
values of 19.3859~MeV and $-38.4422$~MeV  (LDM), 
17.0603~MeV and $-12.8737$~MeV (NLD), or
16.9707~MeV and $-38.9274$~MeV (LSD), depending on the type
of curvature term that is considered (i.e. none at all, a Gaussian one, or one of standard form).
Among the aforementioned models, only the NLD describes fission barriers well, though.
The comparison between the LDM models is complicated by their
different definition of the surface (diffuse in the FRLDM and sharp in 
the models of Ref.~\cite{Pomorski03a}) and the use of different 
shape parametrizations in the study of fission barriers. With the
exception of the LSD, none of these parametrizations comes close to 
the optimal value for a Skyrme EDF, although it has to be noted
that it is not entirely clear how to calculate EDF values for 
$a_{\text{surf}}$ and $a_{\text{ssym}}$ 
that can be meaningfully compared with those of a macroscopic-microscopic 
model (i.e.\ with the HF scheme, or an ETF scheme, or even differently because
of the different density profiles assumed in LDM models). In any event,
all of the above points to the conclusion that one cannot expect
that a parametrization of the nuclear EDF that reproduces the 
$a_{\text{surf}}$ and $a_{\text{ssym}}$ values of a macroscopic-microscopic 
model will perform well for deformation energies.

More relevant experimental data would be most useful to better constrain and benchmark nuclear surface properties; 
in particular data that probe the deformed 
density distribution of heavy nuclei, both for well-deformed ground states of 
heavy nuclei and especially for states at large deformation. The few 
existing measurements of higher-order shape deformations of nuclear 
ground states were all achieved in the 1970s mostly with stable nuclei. Similarly, there is very little information 
available on the excitation energies and quantum numbers of superdeformed states.
In our view, the surface properties of nuclear matter deserve more investment: such information
is as important to fine-tune models as the much-more-often investigated bulk 
properties of nuclear matter.

For the reasons recalled above, among the parametrizations discussed 
in this paper, 1T2T(0.80) is the one that offers the best overall description 
of a wide range of observables. As they all use the full c.m.\ correction energy 
and keep the tensor terms from the two-body central interaction with coupling 
constants $C^{sT}_t$ in Eqs.~\eqref{eq:E:sk:e} and~\eqref{eq:E:sk:o}, 
the EDF of the 1T2T(X) has the same form as the one of SLy7 constructed in 
Ref.~\cite{CHABANAT1998231}. Because of the similarity of the adjustment 
protocol, we encourage the use of 1T2T(0.80) as a replacement of the 
parametrizations of Ref.~\cite{CHABANAT1998231} in future nuclear 
structure studies, and propose that it shall be used under the name of SLy7*.
Its parameters (which can also be found in the supplementary material~\cite{supplement})
are
\begin{center}
\begin{tabular}{lcrlcr}
$t_0$ &=& $-2676.132387$,~~~~ & $x_0$ &=&  $0.574713$, \\
$t_1$ &=&   $381.547873$,~~~~ & $x_1$ &=&  $0.015424$, \\
$t_2$ &=&  $-438.549085$,~~~~ & $x_2$ &=& $-0.892996$, \\
$t_3$ &=& $15893.083082$,~~~~ & $x_3$ &=&  $0.764736$, \\
$W_0$ &=&   $119.182854$,~~~~ & $\alpha$ &=& \multicolumn{1}{l}{$\frac{1}{6}$\,.} \\
\end{tabular}
\end{center}
The parameter $t_0$ is in $\text{MeV}\,\text{fm}^{3}$, $t_1$ and $t_2$ in $\text{MeV}\,\text{fm}^{5}$,
$t_3$ in $\text{MeV}\,\text{fm}^{3+1/6}$, and $W_0$ in $\text{MeV}\,\text{fm}^{5}$. 
The $x_j$ and $\alpha$ are dimensionless.

Unlike SLy4 and SLy6, the form of SLy7* includes
all contributions to the EDF obtained from a two-body generator,
removing some ambiguities about its use in nuclear matter studies. Compared with SLy7 and to the vast 
majority of other parametrizations of the Skyrme EDF at NLO, SLy7* does not 
exhibit finite-size instabilities~\cite{(Pas13a)} in any of the $(S,T)$ channels 
at densities encountered in finite nuclei, such that it can be used for time-reversal
breaking calculations without the need for modifying coupling constants of
the time-odd part of the EDF~\eqref{eq:E:sk:o}. We do not report on such 
calculations here, but we checked the stability of our parametrizations by means of
cranked HFB calculations of rotational bands at high spin~\cite{dacosta_22}.
Results for rotational bands and one-quasiparticle states of odd-mass heavy 
nuclei obtained with time-reversal breaking calculations with SLy7* 
will be reported elsewhere~\cite{Bonnard23x}.

%
%

\begin{acknowledgments}

This project has been supported by the Agence Nationale de la Recherche, France, 
Grant No.~19-CE31-0015-01 (NEWFUN).
W.~Ryssens is a Research Associate of the Fonds de la Recherche Scientifique -- FNRS (Belgium).
The computations were performed using HPC resources from the CC-IN2P3 of the CNRS.

\end{acknowledgments}

%
%

\bibliography{sly7star.bib}

%
%

\end{document}